\documentclass[superscriptaddress,aps,prb,reprint,twocolumn,amsmath,amssymb]{revtex4-2}
\usepackage{graphicx}
\usepackage{hyperref}
\usepackage{amssymb}
\usepackage{slashed}
\usepackage{dcolumn}
\usepackage{amsmath}
\usepackage{bm}% bold math
\usepackage{colordvi}
\usepackage{algorithm}
\usepackage{algpseudocode}
\usepackage{multirow}
\usepackage{newtxtext,newtxmath}
\usepackage{dsfont}
\usepackage{xcolor}
\usepackage{mathbbol}

\usepackage{hyperref}% add hypertext capabilities
\hypersetup{
    colorlinks=true,
    linkcolor=blue,
    citecolor=blue,     
    urlcolor=blue,
}% Added by user
\allowdisplaybreaks% Added by user

\usepackage{mathrsfs}
\makeatletter

\newcommand{\Rmnum}[1]{\expandafter\@slowromancap\romannumeral #1@}
\makeatother

\begin{document}

\preprint{APS/123-QED}

\title{Preformed Cooper pairing and the uncondensed normal-state component in phase-fluctuating {monolayer}  cuprate superconductivity}% Force line breaks with \\

\author{Fei Yang}
 \email{fzy5099@psu.edu}
 \affiliation{%
 Department of Materials Science and Engineering, The Pennsylvania State University, University Park, Pennsylvania 16802, USA
}%

\author{Yin Shi}
 \email{yxs187@psu.edu}
 \affiliation{%
 Department of Materials Science and Engineering, The Pennsylvania State University, University Park, Pennsylvania 16802, USA
}%

\author{Long-Qing Chen}
 \email{lqc3@psu.edu}
 \affiliation{%
 Department of Materials Science and Engineering, The Pennsylvania State University, University Park, Pennsylvania 16802, USA
}%



\date{\today}% It is always \today, today,

\begin{abstract}

We develop a self-consistent microscopic framework  beyond mean-field theory for monolayer cuprate superconductivity. It couples fermionic quasiparticles with collective phase dynamics to treat the gap and superfluid stiffness. The phase sector explicitly incorporates both smooth bosonic Nambu–Goldstone phase fluctuations, renormalized by long-range Coulomb interactions, and topological BKT-type vortex-antivortex fluctuations. The required input is the correlated single-particle spectral function, enabling direct interfacing with  Hubbard-type models. The theory provides access to key superconducting observables, including $T$-dependent gap and  phase stiffness, gap-closing temperature $T_{\rm os}$, and transition temperature $T_c$, across wide ranges of doping. Using a solvable interaction model as input, our simulations reveal several important features
consistent with experimental observations in cuprate superconductors: a $d$-wave superconducting dome in $T$-$p$ phase diagram with a shoulder-like anomaly in underdoped regime, a pronounced separation between $T_c$ and $T_{\rm os}$ signaling preformed Cooper pairing, a finite uncondensed normal component persisting even at $T=0$, and the onset temperature $T_{\rm on,vortex}$ of vortex signals, offering a consistent understanding of how strong correlations and phase fluctuations cooperate to shape high-$T_c$ superconductivity.

\end{abstract}

\maketitle

%\tableofcontents

\section{Introduction}

High-$T_c$ superconductivity in strongly correlated systems has continued to intrigue scientists for nearly four decades~\cite{dagotto94correlated,tom99the,armitage10progress,davis13concepts,damascelli2003angle,keimer15from} since the discovery of cuprate superconductors~\cite{bednorz86possible,wu1987superconductivity}, owing to the coexistence of multiple many-body phases and the complexity of the phase diagrams. In cuprates, doping carriers into the insulating parent compound rapidly suppresses  antiferromagnetic order. 
A $d$-wave superconducting (SC) state sets in at a critical doping, reaches a maximum transition temperature $T_c$ at optimal doping, then quickly declines upon further doping and eventually vanishes~\cite{keimer15from}. The correlated normal state above $T_c$ is equally unconventional. In the overdoped regime, transport deviates strongly from Fermi-liquid behavior, giving rise to the so-called strange metallicity~\cite{martin1990normal,keimer15from}. In the underdoped regime, angle-resolved photoemission spectroscopy (ARPES) directly reveals a finite normal-state gap (pseudogap)~\cite{ding1996spectroscopic,loeser1996excitation,norman1998destruction,valla2006ground}, which persists above $T_c$ and extends up to a much higher temperature. Substantial experimental evidence further points to a variety of intertwined or incipient electronic orders within pseudogap phase~\cite{keimer15from}, including charge-density waves, spin-density waves, and electronic nematicity, which modify the electronic structure and, in principle, compete with the SC phase. 

Notably, unlike conventional bulk superconductors where the Anderson-Higgs mechanism~\cite{anderson63plasmons,ambegaokar61electromagnetic,littlewood81gauge} (3D long-range Coulomb interactions) enforces rigid long-range phase coherence, cuprates are inherently prone to SC phase fluctuations due to their quasi-two-dimensional layered structure~\cite{damascelli2003angle,keimer15from}. Such fluctuations are widely believed to play a pivotal role~\cite{emery1995importance,randeria1992pairing,yang2008emergence,seo2019evidence,benfatto2001phase,yuli2008enhancement,lee2006doping}. A prominent viewpoint holds that the SC transition at $T_c$ in cuprates is not determined by the closing of the pairing gap, but rather by a change in SC phase coherence from long range to short range~\cite{yang2008emergence,seo2019evidence,yuli2008enhancement,lee2006doping,wang2005field,kondo2011disentangling,corson1999vanishing,bilbro2011temporal}. 
This transition occurs at a lower temperature than the gap-closing transition, indicating  that Cooper pairs survive above $T_c$ as preformed but incoherent entities~\cite{yang2008emergence,seo2019evidence,gomes2007visualizing,pasupathy2008electronic,alldredge2008evolution} without the global SC phase coherence necessary for superconductivity (zero resistance).  While thermal fluctuations typically confine such incoherence to a narrow temperature window around $T_c$, experimental advances~\cite{wade1994electronic,mahmood2019locating} have revealed signatures of substantial quantum (zero-point) SC phase fluctuations, leaving a significant fraction of the normal-state component remaining uncondensed and persisting down to $T \rightarrow 0$. Studies using ARPES~\cite{he21superconducting} and relevant theoretical modeling~\cite{li21superconductor} emphasize the role of the presence of a partially flat-band structure near the ($\pi$,0) and (0,$\pi$) regions of the Brillouin zone.  The associated large density of states substantially enhances the pairing susceptibility~\cite{huang19enhanced,sayyad20pairing} against phase fluctuations, promoting local pair formation and granular SC correlations without establishing full global coherence~\cite{li21superconductor}.

For those characteristic features: a $d$-wave SC dome (with a shoulder-like anomaly in underdoped regime), finite uncondensed normal-state component even at $T=0$, and persistence of preformed Cooper pairs above $T_c$, while substantial theoretical progress has been made, several key challenges remain for a fully  comprehensive framework, as capturing phase fluctuations goes beyond the mean-field theory. In general, phase fluctuations comprise both longitudinal {\sl smooth-type}  {bosonic} Nambu–Goldstone (NG) phase fluctuations~\cite{ambegaokar61electromagnetic,nambu1960quasi,nambu2009nobel,littlewood81gauge} and transverse topological Berezinskii–Kosterlitz–Thouless-type (BKT-type) vortex fluctuations~\cite{benfatto2001phase,benfatto10,PhysRevB.110.144518,PhysRevB.80.214506,PhysRevB.77.100506,PhysRevB.87.184505}, which are typically treated separately and in limiting regimes (e.g., NG modes at $T=0$~\cite{PhysRevB.97.054510,PhysRevB.70.214531,PhysRevB.102.060501} and BKT physics near $T_c$~\cite{benfatto2001phase,benfatto10}) or within phenomenological models (such as Gaussian fluctuations in Ginzburg–Landau theory~\cite{PhysRevLett.91.257002} or stochastic {Anderson} disorder in tight-binding model~\cite{li21superconductor}), rather than within a unified, internally consistent microscopic description valid across the full temperature or doping range. State-of-the-art numerical microscopic methods such as dynamical mean-field theory (DMFT), while highly effective for fermionic quasiparticles,  are not yet capable of treating multiple degrees of freedom (fermionic, bosonic and topological) on equal footing and their mutual interplay. Moreover, current theoretical treatments of SC phase fluctuations rely on uncorrelated band models with analytical energy dispersions~\cite{yang21theory,li21superconductor,benfatto2001phase,PhysRevB.97.054510,benfatto10}, whereas in strongly correlated systems the relevant input is the correlated single-particle spectral function~\cite{annurev:/content/journals/10.1146/annurev-conmatphys-090921-033948,PhysRevLett.102.056404,PhysRevX.11.011058,PhysRevB.97.085125,PhysRevX.8.021048,RevModPhys.68.13,vollhardt2011dynamical,avella2013strongly}, typically inferred from Hubbard-type models~\cite{phillips18absence,phillips20exact,yeo19local,zhao22thermodynamics,zhao23failure,zhong22solvable,zhu21topological,tenkila25dynamical,mai23is,mai23topological,setty24electronic,setty24symmetry,PhysRevB.108.235149,li2022two,PhysRevB.108.035121}, whose exact solution is intractable beyond 1D. This presents a limitation in describing phase-fluctuating SC state, particularly when long-range Coulomb interactions, essential for establishing global phase coherence, are also involved in the calculation.

To overcome these limitations, we treat the SC gap and the superfluid density by explicitly incorporating all relevant microscopic ingredients (fermionic quasiparticles, bosonic NG-mode phase dynamics modified by long-range Coulomb interactions, and topological defects of BKT-type  fluctuations). 
The theory is grounded in the correlated single-particle spectral function, ensuring a direct connection with Hubbard-model-based approaches.  This offers a fully self-consistent microscopic framework for phase-fluctuating superconductivity in cuprates, capturing the key SC observables (like gap, phase stiffness, gap-closing temperature $T_{\rm os}$, and SC transition temperature $T_c$) across doping and temperature. As an initial step toward capturing the essential physics, we {focus on monolayer cuprates} and employ a recently developed solvable interaction model~\cite{worm24fermi}.  Remarkably, the simulations reveal several critical features consistent
with experimental observations in cuprate superconductors, including the $d$-wave SC dome in the temperature-doping phase diagram with a shoulder-like anomaly in the underdoped regime, a pronounced separation between $T_c$ and $T_{\rm os}$ signaling preformed Cooper pairing, and a finite uncondensed normal-state component persisting at $T=0^+$,  offering a promising route to understanding the interplay between strong correlations and phase-fluctuating superconductivity in the cuprates.

\section{Phase-Fluctuating SC Model}

{We start from an idealized strictly two-dimensional SC system, appropriate for monolayer cuprate superconductors. Monolayer cuprate superconductors have been realized experimentally over the
past decade~\cite{Jiang2014,Sterpetti2017,Yu2019}, and in particular, experiments have reported that key phenomenology (superconductivity, pseudogap, charge order) of intrinsic monolayer cuprates closely parallels that of bulk compounds (e.g., See Ref.~\cite{Yu2019}), thereby providing a clean and well-defined platform for the physics addressed in the present work.} 

{To facilitate readability, here we provide a concise overview of the effective theoretical framework for two-dimensional SC systems, while the full microscopic derivation  within the fundamental path-integral approach and quantum statistic framework is addressed in Appendix~\ref{appB}. For completeness, possible extension to bulk layered structures and its implications for our main conclusions are briefly discussed in Sec.~\ref{secbulk}.} 
 
Specifically, using Hubbard-Stratonovich transformation, we introduce a generalized SC order parameter $
\Delta({\bm x},{\bm x}')=\sum_{\bm k} e^{i{\bm k}\cdot({\bm{x}-\bm{x}'})}\Delta_{{\bm k}}({\bf R})$,
with the center-of-mass coordinate ${\bm R}=({\bf x+x'})/{2}$~\cite{abrikosov2012methods} and the momentum-dependent component:
\begin{equation} 
\Delta_{{\bm k}}({\bm R}) =|\Delta|\zeta_{\bf k}e^{i\delta\theta({\bm R})}, 
\end{equation}
where the form factor $\zeta_{\bf k}$ specifies the pairing symmetry {(see Appendix~\ref{appA})}, 
\begin{equation} 
\zeta_{\bf k}= \left\{\begin{aligned}
    \cos(k_x)-\cos(k_y), &~~~d_{x^2-y^2}-{\rm wave}\\
    \sin(k_x)\sin(k_y), &~~~d_{xy}-{\rm wave}\\
    1, &~~~s-{\rm wave}
\end{aligned}\right.
\end{equation}
Here the gap amplitude $|\Delta|$ is taken as homogeneous, while phase fluctuations are encoded in $\delta\theta({\bf R})$. The associated phase dynamics ${\bf p}_s={\bf \nabla_R}\delta\theta({\bf R})/2$ naturally decomposes into two orthogonal contributions~\cite{benfatto10}: a longitudinal component ${\bf p}_{s,\parallel}$, describing smooth, long-wavelength NG fluctuations, and a transverse component ${\bf p}_{s,\perp}$ one, associated with BKT-type vortex excitations (topological defects). The NG phase fluctuations~\cite{ambegaokar61electromagnetic,nambu1960quasi,nambu2009nobel,littlewood81gauge} renormalize the SC gap in a gauge-like manner, analogous to the effect of a vector potential~\cite{ambegaokar61electromagnetic,nambu1960quasi,nambu2009nobel}. The resulting SC gap equation within the spectral representation (on-shell approximation) is
\begin{equation}\label{GE}
 \frac{1}{J}\!=\!\bigg\langle\sum_{\bm k}\zeta_{\bf k}^2\int\frac{d\omega}{2\pi}\frac{\tanh[{\beta}E(\omega,{\bf p_{s,\parallel}})/2]}{2\sqrt{\omega^2\!+\!|\Delta|^2\zeta_{\bf k}^2}}A({\bf k},\omega)\bigg\rangle_{\rm NG},
\end{equation}
where $J$ is the pairing strength defined by effective pairing interaction $J_{\bf kk'}=J\zeta_{\bf k}\zeta_{\bf k'}$. The fermionic quasiparticle energy $E(\omega,{\bf p_{s,\parallel}})={\bf v_{\bf k}}\cdot{{\bf p_{s,\parallel}}}\pm\sqrt{\omega^2\!+\!|\Delta|^2\zeta_{\bf k}^2}$, where NG phase fluctuations ${\bf p_{s,\parallel}}$ enter through the Doppler shift~\cite{fulde1964superconductivity,yang2018fulde,yang21theory} and ${\bf v}_{\bf k}$ is the normal-state group velocity. $\beta=1/T$ is the inverse temperature.  The brackets $\langle\dots\rangle_{\rm NG}$ denote statistical averaging over the NG phase fluctuations, while $A({\bf k},\omega)$ represents the correlated single-particle spectral function.

The NG fluctuations arise from the bosonic excitation of the NG mode, a direct consequence of spontaneous $U(1)$ symmetry breaking~\cite{nambu1960quasi,goldstone1961field,goldstone1962broken,nambu2009nobel,schrieffer1964theory,PhysRevB.102.014511}, and the statistical average reads 
\begin{equation}\label{EPF}
\langle{p}^2_{s,\parallel}\rangle=\int\frac{d{\bm q}}{(2\pi)^2}\frac{q^2[2n_B(\omega_{\rm NG})+1]}{2D_{q}\omega_{\rm NG}(q)}. 
\end{equation}
Here, the function $n_B(x)$ is the Bose distribution;  {\small{$\omega_{\rm NG}(q)=\sqrt{2f_sq^2/D_q}$}} is the energy spectrum of the NG mode, as established in the
literature~\cite{sun20collective,yang21theory,ambegaokar61electromagnetic,PhysRevB.64.140506,PhysRevB.69.184510,PhysRevB.70.214531,PhysRevB.97.054510}, where {\small{$D_q=\chi_{\rho\rho}/(1+2\chi_{\rho\rho}V_q)$}}, $\chi_{\rho\rho}$ is the density-density correlation, and {\small{$V_q=2{\pi}e^2/(q\epsilon_0)$}} denotes 2D Coulomb potential. Thus, the gapless NG mode acquires $\omega_{\rm NG}(q)\propto\sqrt{q}$ at long wavelength due to Coulomb interactions, avoiding infrared divergence divergence at $T\ne0$~\cite{yang24thermodynamic,PhysRevB.97.054510}. But the gapless nature indicates active fluctuations, consisting of contributions from both  thermal excitations $2n_B(\omega_{\rm NG})$ and zero-point oscillations. 

The phase stiffness $f_s$ [equivalent to $\hbar^2n_s /(4m^*)$, with $n_s$ the superfluid density and $m^*$ the effective mass] is given by
\begin{equation}\label{fss}
f_s=\Big\langle\!\sum_{\bf k}\!\frac{\mathrm{v}^2_{\bf k}|\Delta|^2\zeta_{\bf k}^2}{1+\xi/l}\!\!\int\!\!\frac{d\omega}{2\pi}\frac{\tanh\big[\beta{E(\omega,{\bf p_{s,\parallel}})}/2\big]}{8(\omega^2\!+\!|\Delta|^2\zeta_{\bf k}^2)^{3/2}}A({\bf k},\omega)\Big\rangle_{\rm NG},
\end{equation}
where the prefactor $(1+\xi/l)^{-1}$ accounts for disorder~\cite{tinkham2004introduction,PhysRevB.106.144509,PhysRevB.99.224511,PhysRevB.98.094507,abrikosov2012methods}, with $\xi$ the coherence length, $l$ the mean free path, and $\xi/l=|\Delta|/\Gamma$ defined through the effective scattering rate $\Gamma$. This bare $f_s$, incorporating both fermionic quasiparticles and NG fluctuations, provides the initial condition for the BKT-type renormalization group equations~\cite{benfatto10,PhysRevB.110.144518,PhysRevB.80.214506,PhysRevB.77.100506,PhysRevB.87.184505},
\begin{equation}
\frac{dK}{dl}=-K^2g^2~~~\text{and}~~~\frac{dg}{dl}=(2-K)g,
\end{equation}
where $K(l=0)={\pi}f_s/T$ and $g(l=0)=2\pi e^{-cK(l=0)}$ with $c=2/\pi$ for 2D or layered systems~\cite{benfatto10}. This yields the renormalized stiffness
\begin{equation}
{\bar f}_s=\frac{T}{\pi}K(l=\infty),
\end{equation}
which further incorporates BKT-type  vortex–antivortex fluctuations. The SC transition temperature $T_c$ follows from $\bar f_s(T_c)=0$, while the gap-closing temperature is denoted $T_{\rm os}$. 

\section{Simulation Results}

To simulate the phase-fluctuating superconductivity in cuprates, the framework requires as input the correlated normal-state single-particle spectral function, interfacing directly with Hubbard-model-based computations.  Such spectral functions typically require advanced numerical many-body calculations, such as DMFT or its diagrammatic extensions. On the other hand, recent studies~\cite{phillips18absence,phillips20exact,yeo19local,zhao22thermodynamics,zhao23failure,zhong22solvable,zhu21topological,tenkila25dynamical,mai23is,mai23topological,setty24electronic,setty24symmetry,PhysRevB.108.235149,li2022two,PhysRevB.108.035121} have explored the solvable Hatsugai–Kohmoto model~\cite{hatsugai92exactly,baskaran91an}, a simplified Hubbard model that yields a Mott insulator at half-filling and a non-Fermi-liquid metal upon doping~\cite{phillips20exact,zhao22thermodynamics}. Accordingly, Ref.~\cite{phillips20exact} examined the $s$-wave SC pairing stability using the Hatsugai–Kohmoto model, based on a mean-field ansatz without phase fluctuations. More recently,  Worm {\em et al.}~\cite{worm24fermi} introduced a simplified interacting model 
\begin{equation}\label{Hamworm}
    \hat{H}_i= \frac{{\mathcal{V}}}{2}\hat{n}_{\bm{k} \sigma} \hat{n}_{\bm{k} + \bm{Q} \bar{\sigma}},~~~~~{\bf Q}=(\pi,\pi),   
\end{equation}
which retains only the interaction component associated with the antiferromagnetic spin fluctuations, yet reproduces essential features of the full  Hubbard model obtained from numerically demanding many-body methods such as the cutting-edge D$\Gamma$A scheme{~\cite{worm23numerical,PhysRevB.75.045118,PhysRevB.80.075104}}, particularly in the doping regime away from the parent Mott insulator, {with particular emphasis on the pseudogap phase}. While this model can not accurately capture the complete physics of cuprates, it provides valuable insights into the related complexity.

\begin{figure}
    \centering
    \includegraphics{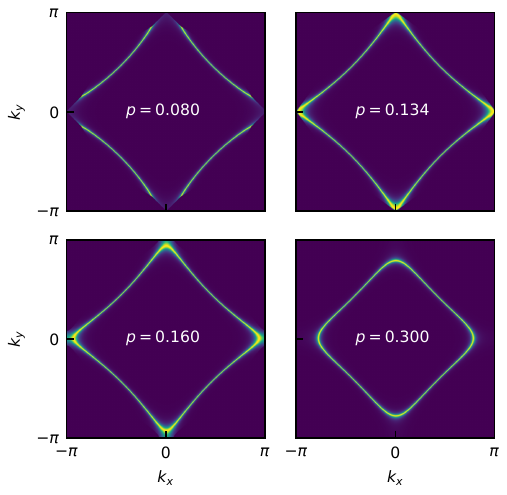}
    \caption{Doping evolution of the Fermi surface from underdoped to overdoped regimes at $T\approx3\,\mathrm{mK}$. We take  ${{\mathcal{V}}}=1.1t$~\cite{worm24fermi}.}
    \label{fig:normal}
\end{figure}

Here we adopt this model as the basis for the initial implementation. From Eq.~(\ref{Hamworm}), the correlated single-particle excitation spectrum reads (see Appendix~\ref{appC})~\cite{worm24fermi} 
\begin{equation}
    A(\bm{k},\omega) = (1 - n_{\bm{k} + \bm{Q}}) \delta(\omega - \epsilon_{\bm{k}}) + n_{\bm{k} + \bm{Q}} \delta(\omega - \epsilon_{\bm{k}} - {{\mathcal{V}}}), \label{eq:A}
\end{equation}
with the electron occupation
\begin{equation}
n_{\bm{k}+\bm{Q}}=\frac{e^{-\beta\epsilon_{{\bm{k}+\bm{Q}}}}+e^{-\beta(\epsilon_{\bm{k}}+\epsilon_{{\bm{k}+\bm{Q}}}+{{\mathcal{V}}})}}{1+e^{-\beta\epsilon_{\bm{k}}}+e^{-\beta\epsilon_{{\bm{k}+\bm{Q}}}}+e^{-\beta(\epsilon_{\bm{k}}+\epsilon_{{\bm{k}+\bm{Q}}}+{{\mathcal{V}}})}}.
\end{equation}
Here ${\bf Q}=(\pi,\pi)$ is the characteristic momentum of the antiferromagnetic-fluctuation–mediated coupling~\cite{worm24fermi} and $\mathcal{V}$ the coupling interaction strength. The underlying uncorrelated single-particle band (shifted by the chemical potential $\mu$) for the square-lattice CuO\textsubscript{2} plane is $
    \epsilon_{\bm{k}} =-2 t [\cos(k_x) + \cos(k_y)] - 4 t' \cos(k_x) \cos(k_y)-\mu$, where the nearest and next-nearest neighbor hoppings are set to typical values $t \approx 0.3\,\mathrm{eV}$ and $t' \approx -0.16 t$,
    {well within the range commonly used in theoretical studies~\cite{nicoletti10high,worm23numerical,phillips20exact,li21superconductor,PhysRevLett.105.077002} and consistent with parametrizations for La-based cuprates such as LSCO and LBCO~\cite{PhysRevB.77.220502,Miao2021}.}     
  This leads to a partially flat band structures  near the ($\pi$,0) and (0,$\pi$) regions of the Brillouin zone. 

{A quantitative comparison between the solvable correlated model and the locally interacting Hubbard model using state-of-the-art many-body techniques was performed in Ref.~\cite{worm24fermi}. Within this mapping, the value ${\mathcal{V}}=1.1t$ in the solvable model was numerically shown to correspond to a Hubbard interaction strength of approximately $U=8t$ in the original model. We adopt this value (${\mathcal{V}}=1.1t$) throughout.}  
 Then, as shown in Fig.~\ref{fig:normal}, this correlated spectral function successfully captures the Fermi-surface evolution from underdoped to overdoped regimes in cuprates. In the underdoped regime the Fermi surface is fragmented into four disconnected Fermi arcs, with no states near $(\pi,0)$ points where a Luttinger surface emerges instead~\cite{worm24fermi,PhysRevB.68.085113}, consistent with ARPES observations of the pseudogap phase in underdoped cuprates~\cite{damascelli2003angle,Vishik_2018}.   Increasing doping above a critical value ($p_c\approx0.13$) triggers the recovery of low-energy spectral weight around $(\pi,0)$, drives the closure of the Fermi surface, and eventually, at $p>0.16$, leads to a change in Fermi-surface topology in which the Fermi surface moves away from $(\pi,0)$. Thus, in the overdoped regime the spectrum resembles that of noninteracting electrons, with Fermi surface largely free from correlation effects.

The simulation results for SC observables are shown in Fig.~\ref{fig:yc1}, using parameters $J=0.9t$ and $\Gamma=0.02t$. As illustrated in Fig.~\ref{fig:yc1}(a) from mean-field calculations without fluctuations, the correlated normal-state electronic structure in cuprates strongly favors the $d_{x^2-y^2}$ pairing symmetry, while significantly suppressing the $s$-wave channel and completely eliminating the $d_{xy}$ channel. The $d$-wave mean-field transition temperature $T_c^{\rm MF}$ exhibits a characteristic dome as a function of doping, reflecting the doping-dependent correlation effects. The dome peaks near $p \approx 0.16$, where the Fermi surface closes around $(\pi,0)$, moves away from this point, and evolves toward a nearly uncorrelated large Fermi surface (Fig.~\ref{fig:normal}). This behavior indicates that the optimal doping is tied to the interplay between correlation effects and Fermi-surface reconstruction.

{With or without SC  phase fluctuations, the $d$-wave superconductivity is suppressed and vanishes below a finite doping level 
$p\approx0.05$ (see Fig.~\ref{fig:yc1}), which is distinct from the Mott critical point at half filling ($p=0$).
 This suppression of superconductivity arises from a rapid collapse of the effective pairing phase space in the pseudogap-like spectral function used as input entering the $d$-wave gap equation. This is primarily a correlation-driven effect in the pseudogap phase within an itinerant description, rather than a consequence of Mottness.} 

Including SC phase fluctuations beyond mean-field theory leads to the following scenario. First, compared to the mean-field values $|\Delta(0)|$ [Fig.~\ref{fig:yc1}(b)] and $T_c^{\rm MF}$ [Fig.~\ref{fig:yc1}(a)], the NG-mode renormalization suppresses the zero-temperature gap $|\Delta(0)|$ [Fig.~\ref{fig:yc1}(b)] and, consequently, the gap-closing temperature $T_{\rm os}$ [red dots in Fig.~\ref{fig:yc1}(d)]. This suppression originates from bosonic zero-point oscillations of the NG mode feeding back into the SC gap equation.  The corresponding experimental manifestations of this effect 
will be discussed later.

\begin{widetext}
  \begin{center}
\begin{figure}[h]
    \includegraphics[width=14.7cm]{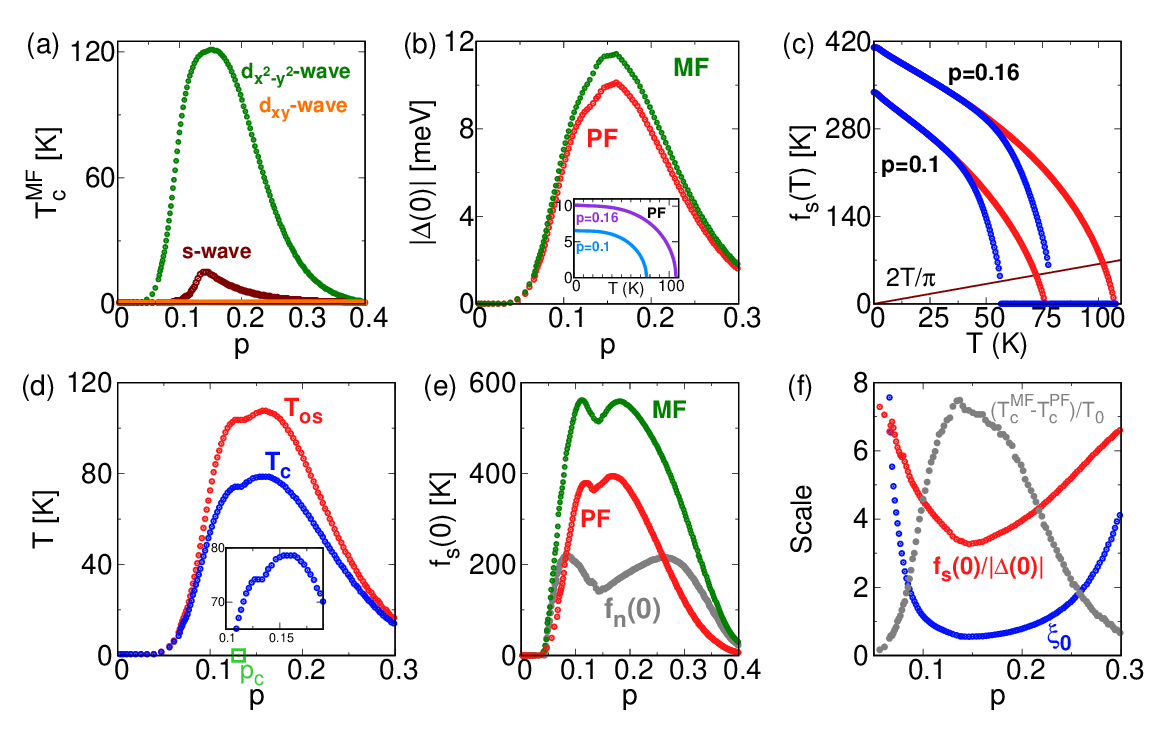}
  \caption{({\bf a}) Transition temperature $T_c^{\rm MF}$ for $d_{x^2-y^2}$-, $d_{xy}$-, and $s$-wave pairings from mean-field theory (without fluctuations).  
({\bf b}) Zero-temperature SC gap within mean-field (MF) theory and the phase-fluctuating (PF) framework. The inset shows $|\Delta(T)|$ in the PF framework.
({\bf c}) Bare (red) and BKT-renormalized (blue) phase stiffness. The straight line indicates $2T/\pi$, where the discontinuous drop of superfluid stiffness occurs.
({\bf d}) Phase diagram of the gap-closing temperature $T_{\rm os}$ and transition temperature $T_c$ as a function of doping within the PF framework. The inset magnifies the shoulder-like feature on the $T_c$-$p$ dome.
({\bf e}) Zero-temperature superfluid stiffness $f_s(0)$, reflecting $n_s(0)/m^*$, within MF and the PF framework, and normal-state component $f_n(0)$.
({\bf f}) Ratio $f_s(0)/|\Delta(0)|$, coherence length $\xi_0$ (normalized by $\xi_0(p=0.15)$) and temperature difference 
$T_c^{\rm MF}-T_c^{\rm PF}$ (normalized by $T_0=6~$K).}    
\label{fig:yc1}
\end{figure}
\end{center}
\end{widetext}

In contrast, the thermal contribution of NG fluctuations to SC observables remains minimal, as the NG-mode thermal  excitations $n_B(\omega_{\rm NG})$ are strongly suppressed. This suppression arises from the modification of the NG dispersion from $\omega_{\rm NG}(q){\propto}q$ to $\omega_{\rm NG}(q)\propto\sqrt{q}$ by long-range Coulomb interactions, which confine low-energy excitations to a narrow momentum window with limited phase space. As a result, these modes propagate rapidly and are only weakly thermally populated,  rendering their effect on thermodynamics and critical dynamics negligible. Consequently, even though the SC state in the cuprates is subject to significant phase fluctuations (NG-mode zero-point oscillation and BKT-type vortex fluctuations), the temperature dependence of the SC-gap amplitude $|\Delta(T)|$ [inset of Fig.~\ref{fig:yc1}(b)] essentially follows the mean-field form  and can be well described by BCS behavior of a $d$-wave gap, consistent with the experimental observations of the gap behavior from ARPES and tunneling spectroscopy~\cite{Lee_2009,Lee2007,Hashimoto2014}.

Including BKT-type vortex fluctuations at elevated temperatures cause the renormalized phase stiffness ${\bar f_s}(T)$ [blue dots in Fig.~\ref{fig:yc1}(c)] to decrease more rapidly than its bare value $f_s(T)$ [red dots in Fig.~\ref{fig:yc1}(c)], ultimately leading to a discontinuous drop of the superfluid density and driving the SC transition at $T_c$. As a result, $T_c$ lies below the gap-closing temperature $T_{\rm os}$ [Fig.~\ref{fig:yc1}(d)]. This separation gives rise to a SC-pseudogap (incoherent preformed Cooper pairing) regime, where the pairing persists without the global SC phase coherence and hence without zero resistivity, consistent with the phenomenology widely observed in cuprates~\cite{Pelc2018,yang2008emergence,seo2019evidence,gomes2007visualizing,pasupathy2008electronic,alldredge2008evolution}. This persisting SC gap above $T_c$ from our simulation also agrees with the observation of the collective Higgs mode  above the SC  transition, as this mode corresponds to the dynamic amplitude fluctuations $\delta|\Delta(t)|$ of the SC  gap~\cite{Chu2020,PhysRevLett.120.117001,PhysRevB.102.054510}, thereby providing direct experimental evidence for the persistence of SC gap and hence pairing above SC transition temperature $T_c$.

Overall, the phase fluctuations (i.e., combined effect of NG zero-point fluctuations and BKT-type fluctuations) produce a plateau (shoulder-like feature) in the rising part of the $d$-wave SC dome of $T_c$ around $p_c\approx0.13$ in the underdoped regime, as illustrated in Fig.~\ref{fig:yc1}(d). This anomaly originates from the recovery of low-energy spectral weight near $(\pi,0)$; consequently, the Fermi surface closes and becomes tangent to the partially flat band near $(\pi,0)$, leading to a sudden enhancement of the effective mass. Such abrupt effective-mass enhancement is consistent with experimental measurements in cuprates~\cite{PhysRevResearch.3.043125,PhysRevB.106.195110,annurev:/content/journals/10.1146/annurev-conmatphys-030212-184305} and, correspondingly, results in a pronounced suppression of the phase stiffness $f_s(0)$ around $p_c$ [Fig.~\ref{fig:yc1}(e)], 
 thereby locally depressing $T_c$ on the $d$-wave SC dome. Notably, most measured $T_c$–$p$ phase diagrams of cuprates~\cite{keimer15from,dagotto94correlated,tom99the,armitage10progress,davis13concepts,damascelli2003angle} exhibit a similar shoulder-like anomaly in the underdoped regime, which has been attributed to fluctuation effects~\cite{keimer15from}. Moreover, in contrast to approaches employing parabolic dispersion~\cite{yang21theory} or artificially steep band structures near $(\pi,0)$~\cite{li21superconductor}, we find that the presence of a partially flat band in the normal-state dispersion around $(\pi,0)$ amplifies the effect of phase fluctuations, thereby making the shoulder-like anomaly more pronounced. This occurs because the partially flat band suppresses the phase stiffness $f_s$~\cite{li21superconductor,he21superconducting}. This, in turn, lowers both the energy of the NG mode, $\omega_{\rm NG} \propto \sqrt{f_sq}$, and the bare coupling constant $K(l=0)$,  
  and hence, amplifies the effect of phase fluctuations.

{It has been suggested that correlation-driven changes in the normal-state spectral function, most notably the recovery of low-energy spectral weight, are linked to various anomalous behaviors in the underdoped cuprates.
In the present context, this mechanism manifests as a pronounced suppression of the phase stiffness, giving rise to a shoulder-like feature in $T_c$ around $p\approx0.13$. Our simulation identifies this feature as occurring on the underdoped side of optimal doping and views it as a generic consequence of correlation-driven spectral-weight recovery. In this sense, it is a primarily \emph{correlation-driven} effect.}

As illustrated in Fig.~\ref{fig:yc1}(e), compared to the mean-field theory without phase fluctuations, the zero-temperature superfluid stiffness $f_s(0)$ is significantly suppressed due to NG-mode zero-point oscillations. This indicates that a substantial fraction of the normal-state component remains uncondensed even at $T\to0^+$, so that the residual normal-fluid contribution is
\begin{equation}
f_n(0)=f_s^{\rm MF}(0)-f_s^{\rm PF}(0),
\end{equation}
or equivalently $\frac{n_n}{m^*}=\frac{n_s^{\rm MF}-n_s^{\rm PF}}{m^*}$, 
with $n_n$ the density of residual normal carriers arising from zero-point NG SC phase fluctuations. This finite normal-state component at $T=0$ is consistent with the observed two-fluid behavior in low-temperature optical conductivity measurements of the cuprates~\cite{mahmood2019locating}, 
\begin{equation}
\sigma(\omega, T=0^+)=\frac{e^2f_s(0^+)}{i\omega}+\frac{e^2f_n(0^+)}{i\omega+\tau^{-1}},
\end{equation}
which concluded that a finite residual normal-fluid component persists uncondensed even at low-temperature limit $T\to0^+$.

The ratio of SC phase stiffness to the gap at zero temperature, $f_s(0)/|\Delta(0)|$, shown in Fig.~\ref{fig:yc1}(f),  is minimized near optimal doping, while it is significantly larger in both the underdoped and overdoped regimes. Consequently, the impact of phase fluctuations, i.e., the temperature difference $T_c^{\rm MF}-T_c^{\rm PF}$ [gray dots in Fig.~\ref{fig:yc1}(f)] or the observable temperature difference $T_{\rm os}-T_c$ signaling preformed Cooper pairing [Fig.~\ref{fig:yc1}(d)], is strongest around optimal doping and becomes less pronounced away from it, exactly consistent with the experimental observation in Higgs-mode optical spectroscopy of  cuprates~\cite{Chu2020,PhysRevLett.120.117001,PhysRevB.102.054510}.

In the preformed Cooper pairing regime above $T_c$, the vortex–antivortex pairs unbind, leading to a resistive state. Applying an external magnetic field $B$ then induces additional free vortices that can be detected in transport and magnetization measurements. This defines the onset temperature of vortex signals, $T_{\rm on,vortex}$, above which vortex-related responses (such as a finite Nernst signal or vortex Hall effect with $\rho_{xy}\ne0$) first become discernible under an applied field. In general, $T_{\rm on,vortex}>T_c$, since a finite density of induced free vortices is required to generate a detectable signal. Consequently, below $T_c$, both $\rho_{xx}=0$ and $\rho_{xy}=0$ (true SC state); within the intermediate window $T_c<T<T_{\rm on,vortex}$, resistive vortex flow is present while the Hall response remains negligible ($\rho_{xx}\neq0$, $\rho_{xy}\approx0$); and above $T_{\rm on,vortex}$, both $\rho_{xx}\neq 0$ and $\rho_{xy}\neq 0$ (phase-incoherent pairing state with a finite Hall signal).

In BKT physics, 
 the thermally generated free-vortex density is given by~\cite{benfatto10,PhysRevB.110.144518,PhysRevB.80.214506,PhysRevB.77.100506,PhysRevB.87.184505}
\begin{equation}
n^{\rm th}_{v}\approx\frac{1}{\xi_0^2}\exp(-b/\sqrt{t}),
\end{equation}
 where $t=(T-T_c)/T_c$, $b$ is a dimensionless constant in the BKT physics and $\xi_0=\hbar{\mathrm{v}_F}/|\Delta(0)|$ the coherence length.  Applying an external magnetic field $B$ can generate the net vortex density $n^B_{v}=B/\Phi_0$ with $\Phi_0$ being the flux quantum. 
 
 Consequently, when the net vortex density becomes appreciable [i.e., the field-induced vortex density $n_v^{B}=B/\Phi_0$ becomes comparable to the thermally generated free-vortex density $n_v^{\rm th}(T_{\rm on,vortex})$], one has
\begin{equation}
\delta{\bar T}=\frac{T_{\rm on,vortex}-T_c}{T_c}\propto\Big[\frac{1}{\ln(\Phi_0/(B\xi_0^2))}\Big]^2.\label{FFF}
\end{equation}
leading to the onset temperature of vortex signals, $T_{\rm on,vortex}$, above which vortex-related responses (such as a finite Nernst signal or vortex Hall effect with $\rho_{xy}\ne0$) first become discernible under an applied field.

The ratio $\delta{\bar T}$ in Eq. (\ref{FFF}) grows with increasing $\xi_{0}$, which is minimized near optimal doping and grows substantially in both the underdoped and overdoped regimes [Fig.~\ref{fig:yc1}(f)]. Consequently, the window $T_{\rm on,vortex}-T_c$ or the ratio $(T_{\rm on,vortex}-T_c)/T_c$ remains small and confined near optimal doping but broadens substantially toward both the underdoped and overdoped sides of the dome, consistent with experimental finding in Ref.~\cite{terzic2024persistence}.

\section{Discussion}
\label{secbulk}

{Despite the fact that the above formulation and numerical analysis focus on monolayer cuprates, the framework can be straightforwardly extended to bulk layered cuprate superconductors.   Bulk cuprates are quasi-two-dimensional rather than strictly two-dimensional systems. Their intrinsic three-dimensional layered structure gives rise to long-range Coulomb interactions that qualitatively modify the collective phase dynamics compared with an idealized strictly two-dimensional limit. Below, we clarify how the present phase-fluctuation framework can be naturally extended to bulk layered systems, and discuss the resulting physical implications.}

{Specifically, for an evenly spaced layered superconductor, one can follow the derivation outlined in Sec.~\ref{appA}. After integrating out the fermionic degrees of freedom and retaining only the SC phase degrees of freedom, one arrives at a layer-resolved phase-only action of the form 
\begin{align}
\!S\!=\!&\sum_n\!\int\!\!{dR}\Bigg\{\frac{\chi_{\rho\rho}}{2}\Big[\frac{\partial_t\delta\theta_n(R)}{2}\!+\!\mu_{H}(R,n)\Big]^2\!-\!\frac{f_s}{2}\Big[\frac{\nabla\delta\theta_n(R)}{2}\Big]^2\nonumber\\
&\mbox{}+\!J_c\cos(\theta_{n+1}\!-\!\theta_n)\Bigg\}\!+\!\frac{1}{2}\int\!\!{dt}{d{\bf q}_{\parallel}}dq_z\frac{|\mu_{\rm H}({\bf q}_{\parallel},q_z)|^2}{V({\bf q}_{\parallel},q_z)},    
\end{align}
where $\delta\theta_n(R)$ and $\mu_H(R,n)$ denote the SC phase fluctuation and the Hartree field in the $n$-th layer, respectively, and $J_c$ is the interlayer Josephson coupling energy density. Here, ${\bf q}_{\parallel}$ denotes the in-plane momentum and $q_z$ the out-of-plane lattice momentum. Owing to the discrete layered structure, the phase-fluctuation field and the Hartree field can be expressed in momentum space as
\begin{eqnarray}
\delta\theta_n(R)&=&\sum_{{\bf q}_{\parallel},q_z}\delta\theta({\bf q}_{\parallel},q_z)e^{i({\bf q}_{\parallel}\cdot{\bf R}+q_znd)},\\
\mu_{H}(R,n)&=&\sum_{{\bf q}_{\parallel},q_z}\mu_{H}({\bf q}_{\parallel},q_z)e^{i({\bf q}_{\parallel}\cdot{\bf R}+q_znd)},     
\end{eqnarray}
with $d$ the layer spacing. {Here $V({\bf q}_{\parallel},q_z)$ denotes the long-range Coulomb interaction in momentum space. 
In cuprates, the appropriate momentum-space form of the long-range Coulomb interaction is subtle and remains under active discussion in the literature, particularly when interlayer structure (e.g., in superconductors with more than one layer per unit cell) and microscopic screening effects are taken into account~\cite{PhysRevB.102.024509,PhysRevB.111.085138,PhysRevB.109.144516,PhysRevB.111.104509}. As a commonly used representative example, for single-layer bulk cuprates such as LSCO,  one may adopt~\cite{PhysRevB.102.024509} 
\begin{equation}
 V({\bf q}_{\parallel},q_z)=\frac{e^2a^2/(2d)}{{\varepsilon_{\parallel}}(2-\cos{q_x}a-\cos{q_y}a)+{\varepsilon_{\perp}}(1-\cos{q_zd})}, 
\end{equation}
where $\varepsilon_{\parallel}$ and $\varepsilon_{\perp}$  are the dielectric constants parallel and perpendicular to the planes, respectively.}

{Within this description, the collective phase-mode spectrum is then derived as
\begin{equation}
\omega^2({\bf q}_{\parallel},q_z)=\frac{1}{2D_{\bf q}}\Big[4f_sq^2_{\parallel}+2J_c(1-\cos{q_zd})\Big].   
\end{equation}
where $D_{\bf q}=\chi_{\rho\rho}/(1+2V_{\bf q}\chi_{\rho\rho})$. Consequently, for $q_zd=0$, corresponding to in-phase oscillations across layers, the long-wavelength limit of this dispersion, $\omega^2({\bf q}_{\parallel},q_z=0)=\frac{e^2}{d\varepsilon_{\parallel}}f_s+\frac{2f_sq^2_{\parallel}}{\chi_{\rho\rho}}$, describes the in-plane plasmon mode~\cite{PhysRevB.102.024509,sun20collective}, whose characteristic energy scale of excitation gap is pushed to high values, $\sqrt{\frac{e^2}{d\varepsilon_{\parallel}}f_s}$, typically on the order of eV (at low temperatures). In contrast, for finite $q_zd\ne0$, corresponding to relative phase fluctuations between neighboring layers, one obtains out-of-phase collective phase modes, namely the Josephson plasma modes~\cite{sun20collective}, whose characteristic energy scale of excitation gap is set by the interlayer Josephson coupling and typically lies in the meV range at $q_zd=\pi$, as confirmed by experimental measurements~\cite{Dienst2013}.}

{Importantly, within our framework, the influence of phase fluctuations on  superconductivity (fermionic observables) in the CuO$_2$ planes arises through the in-plane superflow generated by spatial phase gradients, entering the quasiparticle spectrum via the Doppler-shift mechanism. The relevant fluctuation strength is controlled by averages of the in-plane phase gradients $\langle{p_s^2}\rangle=\langle{q^2_{\parallel}\delta\theta({\bf q}_{\parallel},q_z)/4}\rangle$.   Therefore, in principle, the present framework can be naturally extended to bulk layered cuprates by directly incorporating the full phase-mode spectrum $\omega({\bf q}_{\parallel},q_z)$ into the analysis  [i.e., into Eq.~(\ref{EPF}) in the main text], yielding
\begin{equation}
\langle{p}^2_{s,\parallel}\rangle=\int\frac{d{\bf q}_{\parallel}dq_z}{(2\pi)^3}\frac{q^2}{D_{\bf q}}\frac{2n_B[\omega({\bf q}_{\parallel},q_z)]+1}{2\omega({\bf q}_{\parallel},q_z)}. 
\end{equation}
Then, only collective phase modes with sufficiently low energies contribute appreciably to these phase fluctuations. As a result, the high-energy in-plane plasmon modes at $q_z=0$ do not contribute significantly, whereas the low-energy collective phase modes at finite $q_z$, whose characteristic energy scale lies in the meV range, provide the dominant contribution to phase fluctuations, including zero-point/quantum oscillations (associated with the vacuum contribution, $+1$ terms) and thermal fluctuations (governed by the Bose occupation factor $2n_B(\omega)$).}

{We also note that the BKT physics is governed by transverse vortex excitations, which do not directly couple to charge density fluctuations, with Coulomb interactions entering only indirectly through a renormalization of the superfluid stiffness [Eqs.~(\ref{EPF}) and~(\ref{fss})].
 By contrast, a weak interlayer Josephson coupling introduces an infrared cutoff to the strictly two-dimensional BKT physics. Over intermediate length scales, the system exhibits BKT-like vortex–antivortex physics within each CuO$_2$ plane, while at the longest length scales the interlayer coupling locks the phases of neighboring layers and the transition crosses over to an anisotropic 3D-XY behavior~\cite{FISCHER1993179,PhysRevB.45.13076,PhysRevB.96.214512,PhysRevB.45.12632}. Practically, this effect can be incorporated by performing the standard two-dimensional BKT renormalization-group flow for the in-plane stiffness $f_s$ and vortex fugacity up to a maximum scale $l_{\rm max}\sim\frac{1}{2}\ln[f_s/(4J_cd^2)]$, rather than taking the two-dimensional limit $l\to\infty$.} 
 
{A fully self-consistent treatment of these extensions would substantially increase the complexity of the analysis at the present stage, as it would require a systematic exploration of the out-of-plane momentum dependence of phase fluctuations and their feedback on SC observables. We leave a fully self-consistent
implementation for layered structures to future work. Physically, as demonstrated above, in layered systems, the relevant low-energy phase modes at finite $q_z$ acquire a small Josephson-plasma gap (meV scale), which primarily regulates the infrared behavior compared with the strictly two-dimensional case. It is expected not to  qualitatively alter the dominance of zero-point fluctuations at low temperature, but this gap and the relatively large group velocity of these collective modes would significantly reduce the available phase space in the bosonic thermal excitation integrals, thereby suppressing thermal phase fluctuations. We thus expect quantitative changes upon this extension, whereas our central physical conclusions remain qualitatively unchanged, in particular the essential role of zero-point NG phase fluctuations and the minimal contribution from thermal NG phase fluctuations emphasized in the present work.}

\section{Summary}

The phase-fluctuating SC theory captures the fact that, in many high-$T_c$ superconductors, the SC phase fluctuates~\cite{rullier11high,emery1995importance,randeria1992pairing,yang2008emergence,seo2019evidence,benfatto2001phase,yuli2008enhancement,lee2006doping,wang2005field,kondo2011disentangling,corson1999vanishing,bilbro2011temporal,wade1994electronic,mahmood2019locating,he21superconducting,li21superconductor}, and these fluctuations can strongly influence SC observables beyond mean-field description, i.e., the high-$T_c$ superconductivity is driven not only by Cooper pair formation but also by long-range phase dynamics.  In cuprates, strong  correlations and nodal $d$-wave pairing amplify the effect of phase fluctuations. Our framework that
incorporate all relevant microscopic ingredients into a single, self-consistent description under the interplay with strong correlations, naturally describes the finite uncondensed  normal-fluid component at $T=0$, and the persistence of preformed Cooper pairs above $T_c$ that lack long-range SC phase coherence and result in a  incoherent-pairing (pseudogap) regime,  successfully gives rise to a $d$-wave SC dome in much closer agreement with experiments of cuprate superconductors, and reproduces various  experimental observations of the nonequilibrium/optical properties across doping and temperature. The framework is designed with an input of the single-particle spectral function, offering an efficient method for directly interfacing with other Hubbard-model–based
computations to study cuprates and other high-$T_c$ superconductors.

{\sl Acknowledgments.---}This work is supported by the US Department of Energy, Office of Science, Basic Energy Sciences, under Award Number DE-SC0020145 as part of Computational Materials Sciences Program. F.Y. and L.Q.C. also appreciate the generous support from the Donald W. Hamer Foundation through a Hamer Professorship at Penn State.

\begin{appendix}

\begin{widetext}

\section{Pairing interaction}
\label{appA}
In this section, we introduce the pairing interaction that gives rise to the spin-singlet pairing potential in the main text.  The microscopic origin of the pairing interaction in strongly correlated cuprates remains debated, with several possible mechanisms having been proposed~\cite{miyake2007new,10.1063/1.2199437,Tranquada01102020,RevModPhys.75.913}.  A  distinct but essential  contribution arises from antiferromagnetic spin fluctuations~\cite{miyake2007new,10.1063/1.2199437,RevModPhys.75.913}, where the low-energy spin susceptibility provides the dominant pairing channel. Using simple Lindhard-type expression for   susceptibility, however, is inadequate in cuprates, as it fails to capture key experimental features such as the paramagnon spectrum observed by inelastic neutron and resonant inelastic x-ray scattering~\cite{PhysRevB.104.L020510,PhysRevB.110.224431}. This contribution is more realistically described within fluctuation-exchange~\cite{PhysRevB.55.2122,PhysRevLett.97.067001,PhysRevB.69.165113} or spin-fermion~\cite{Abanov01032003} approaches, reflecting a strongly renormalized, momentum-dependent coupling shaped by strong correlations, collective spin/charge dynamics, and lattice geometry~\cite{SPALEK20221}. Since our aim at this stage is to establish a tractable framework that captures the essential physics of superconducting phase fluctuations, while treating the pairing interaction as an input,  we consider a generalized pairing interaction in real space~\cite{li21superconductor}:
\begin{eqnarray}
    \hat{H}_\mathrm{ex}&=&   \sum_{i,j}{J_{ij}}\hat{c}_{i \uparrow}^\dagger \hat{c}_{i \downarrow} \hat{c}_{j \downarrow}^\dagger \hat{c}_{j \uparrow},  
\end{eqnarray}
which in momentum space takes the form: 
\begin{equation}
H_{\rm ex}=-\sum_{\bf{kk'}}J_{\bf kk'}\hat{c}^\dagger_{\bf{k}\uparrow} \hat{c}^\dagger_{\bf{-k}\downarrow} \hat{c}_{\bf{-k'}\downarrow} \hat{c}_{\bf{k'}\uparrow},
\end{equation}  
with $J_{ij}>0$ being the interaction strength and $\hat{c}_{i\sigma}$ ($\sigma=\uparrow,\downarrow$) the annihilation operator for  electron with spin $\sigma$ at site $i$. For on-site, nearest-neighbor, and next-nearest-neighbor  couplings on a square lattice, the corresponding momentum-space kernels read
$J^{(0)}_{\mathbf{k}\mathbf{k}'} =J_0$, $J^{(1)}_{\mathbf{k}\mathbf{k}'}= \tfrac{J_1}{2}[\cos(k_x-k_x')+\cos(k_y-k_y')]$, $J^{(2)}_{\mathbf{k}\mathbf{k}'}=2J_2\cos(k_x-k_x')\cos(k_y-k_y')$, respectively. Neglecting the odd-parity parts that do not contribute to the spin-singlet pairings, one has 
\begin{eqnarray}
J_{\bf kk'}&=&\big[J_0+\frac{J_1}{4}(\cos{k_x}+\cos{k_y})(\cos{k_x'}+\cos{k_y'})+2J_2\cos{k_x}\cos{k_y}\cos{k_x'}\cos{k_y'}\big]+\frac{J_1}{4}(\cos{k_x}-\cos{k_y})(\cos{k_x'}-\cos{k_y'})\nonumber\\
&&\mbox{}+2J_2\sin{k_x}\sin{k_y}\sin{k_x'}\sin{k_y'}.
\end{eqnarray}
Keeping only the dominant channels
 and projecting onto separable forms that make the pairing symmetry explicit,  one finds
\begin{equation}
J_{\bf kk'}=J\zeta_{\bf k}\zeta_{\bf k'}, \label{JKK}
\end{equation}
where the form factor $\zeta_{\bf k}$ specifies the pairing symmetry, 
\begin{equation} 
\zeta_{\bf k}= \left\{\begin{aligned}
    \cos(k_x)-\cos(k_y), &~~~d_{x^2-y^2}-{\rm wave}\\
    \sin(k_x)\sin(k_y), &~~~d_{xy}-{\rm wave}\\
    1, &~~~s-{\rm wave}
\end{aligned}\right.
\end{equation} 
Here, to enable a general comparison between different pairing symmetries, we consider the case where the $s$-wave, $d_{x^2-y^2}$-wave and $d_{xy}$-wave channels are assigned the same pairing potential $J$. This choice allows us to compare the relative stability of different pairing symmetries under identical interaction conditions, and highlight the role of correlated normal-state electronic structure in determining the high-$T_c$ superconductivity.

In our calculation, the $d$-wave superconducting dome in the $T$–$p$ phase diagram does not rapidly disappear once the doping level exceeds the optimal value; rather, it extends into the overdoped regime with a residual tail.  This behavior originates from our initial assumption of a constant pairing potential $J$. In reality, once the Fermi surface shifts away from the antiferromagnetic zone boundary, where scattering from antiferromagnetic spin fluctuations is strongest, the effective pairing interaction is expected to drop rapidly~\cite{RevModPhys.84.1383}, giving rise to the experimentally observed abrupt suppression of superconductivity in the overdoped regime. Incorporating a realistic doping dependence $J(p)$ would remove the tail, but this refinement is beyond the scope of the present study and we expect it does not affect the main conclusions of our work.

\section{Derivation of self-consistent phase-transition theory of phase-fluctuating superconductivity}
\label{appB}

The self-consistent phase-transition theory for phase-fluctuating $d$-wave superconductivity was previously developed under the assumption of a parabolic normal-state dispersion~\cite{yang21theory}, primarily within the gauge-invariant kinetic-equation framework~\cite{yang19gauge}, with  a brief path-integral treatment, and restricted to long-wavelength (smooth-type) Nambu–Goldstone phase fluctuations. For completeness, in this section, we  present a comprehensive path-integral derivation that accommodates a generalized normal-state dispersion and further incorporates topological Berezinskii–Kosterlitz–Thouless (BKT)-type fluctuations.

\subsection{Derivation of effective action}

We start with a generalized action of superconductors~\cite{schrieffer1964theory,yang21theory,yang24thermodynamic}: 
\begin{eqnarray}
  S[\psi,\psi^{\dagger}]&=&\int{dx}\sum_{s}\psi^{\dagger}_s(x)[i\partial_{x_0}-\xi_{\hat {\bf p}}]\psi_s(x)+\int{dx}{dx'}J(x,x')\psi^{\dagger}_{\uparrow}(x)\psi^{\dagger}_{\downarrow}(x')\psi_{\downarrow}(x')\psi_{\uparrow}(x)\nonumber\\
  &&\mbox{}-\frac{1}{2}\sum_{ss'}\int{dxdx'}V(x-x')\psi_s^{\dagger}(x)\psi_{s'}^{\dagger}(x')\psi_{s'}(x')\psi_s(x),
\end{eqnarray}
where $\psi_s(x)$ is the electron field operator with the four-vector ${x}=(x_0,{\bf x})$ and spin $s$; $\xi_{\hat {\bf p}}$ denotes a generalized normal-state energy dispersion with ${\hat {\bf p}}=-i\hbar{\nabla}$ being the momentum operator; $J(x,x')$ is a generalized pairing potential and $V(x-x')$ denotes the Coulomb potential in the space-time coordinate. The Fourier component $J_{\bf kk'}$ of the pairing potential $J(x,x')$ is expressed as Eq.~(\ref{JKK}). Then, using the Hubbard-Stratonovich transformation~\cite{schrieffer1964theory,yang21theory,sun20collective}, the action becomes:
\begin{eqnarray}
S[\psi,\psi^{\dagger}]\!=\int\!\!dx{dx'}\Big\{\!\sum_{s}\psi^{\dagger}_s(x)[i\partial_{x_0}\!-\!\xi_{\hat {\bf p}}\!-\!\mu_H(R)]\psi_s(x)\!-\!\Psi^{\dagger}(x){\hat \Delta}(x,x')\Psi(x')\Big\}\!-\!\int\!\!{dR}\frac{|\Delta|^2}{J}\!+\!\frac{1}{2}\int\!\!{dt}d{\bf q}\frac{|\mu_H(t,{\bf q})|^2}{V_{\bf q}}.~~~~~
\end{eqnarray}
Here, $\Psi(x)=[\psi_{\uparrow}(x),\psi^{\dagger}_{\downarrow}(x)]^T$ represents the field operator in the Nambu space; the four-vector  ${R}={(x+x')}/{2}=(t,{\bf R})$ denotes the center-of-mass coordinate of the pairing electrons; $\mu_H(R)$ stands for the Hartree field due to the long-range Coulomb interactions, while $\mu(t,{\bf q})$ and $V_{\bf q}$ are the corresponding Fourier components of Hartree field and Coulomb potential, respectively; the pairing matrix is given by: 
\begin{equation}
{\hat \Delta}(x,x')=\Delta(x,x')\tau_++\Delta^*(x,x')\tau_-,
\end{equation}
with $\tau_i$ being the Pauli matrices in Nambu space~\cite{abrikosov2012methods}. The generalized  superconducting order parameter takes the form:
\begin{equation}
\Delta(x,x')=\sum_{\bm k} e^{i{\bm k}\cdot({\bm{x}-\bm{x}'})}\Delta_{{\bm k}}({R}),
\end{equation}
with the specific form:  
\begin{equation}
\Delta_{{\bm k}}({\bm R}) =|\Delta|\zeta_{\bf k}e^{i\delta\theta({\bm R})}.
\end{equation}
where the gap amplitude $|\Delta|$ is taken as homogeneous, while phase fluctuations are encoded in $\delta\theta({\bf R})$. Applying the unitary transformation: 
\begin{equation}
\Psi(x){\rightarrow}e^{i\tau_3\delta\theta(R)/2}\Psi(x),
\end{equation}
to effectively remove the superconducting phase from the order parameter, the action becomes
\begin{equation}
S\!=\!\!\int\!{dx}{dx'}\Big\{\!\sum_{s}\psi^{\dagger}_s(x)\big[i\partial_{x_0}\!-\!\frac{\partial_{t}\delta\theta(R)}{2}-\xi_{{\hat {\bf p}}+{\bf p}_s}-\mu_H\big]\psi_s(x)-\Psi^{\dagger}(x)|{\Delta}(x,x')|\tau_1\Psi(x')\Big\}-\!\!\int\!\!{dR}\frac{|\Delta|^2}{J}+\int\!\!\!{dt}d{\bf q}\frac{|\mu_H(t,{\bf q})|^2}{2V_{\bf q}},
\end{equation}
where the associated phase dynamics ${\bf p}_s={\bf \nabla_R}\delta\theta({\bf R})/2$.

 Then, by separating the center-of-mass $[R=(t,{\bf R})]$ and relative $[r=x_1-x_2=(\tau,{\bf r})]$ coordinates of the pairing electrons~\cite{schrieffer1964theory,yang21theory}, using the relation $\xi_{{\hat {\bf p}}+{\bf p}_s}\approx\xi_{\bf {k}}+{\bf p}_s\cdot{\bf v_{k}}+\eta{\bf p}_s^2$ with ${\bf v_{k}}=\partial_{\bm k}\xi_{\bm k}$ being the group velocity and $\eta=\partial_{\bm k}^2\xi_{\bm k}/2$, the action simplifies:
\begin{eqnarray}
  S=\int\!{dR}\Big\{\sum_{\bf k}\Psi^{\dagger}_{\bf k}\Big[i\partial_{\tau}\!-\!{\bf p}_s\cdot{{\bf v}_{\bf {k}}}\!-\!\xi_{{\bf k}}\tau_3\!-\!|\Delta_{{\bf k}}|\tau_1\!+\!\Big({{\eta}p_s^2}\!+\!\mu_H\!+\!\frac{\partial_t\delta\theta}{2}\Big)\tau_3\Big]\Psi_{\bf k}\!-\!\Big[{\sum_{\bf k}({\eta}p_s^2}\!+\!\mu_H)\!+\!\frac{|\Delta|^2}{J}\Big]\Big\}\!+\!\int{dt}d{\bf q}\frac{|\mu_H(t,{\bf q})|^2}{2V_q}.~~~
\end{eqnarray}
Here, ${{\bf k}}$ denotes the relative-momentum operator. Additionally, using the fact $V_{\bf q=0}\equiv0$,  the integral $\int{dR}\mu_H(R)=0$. 

After performing the standard integration over the Fermi field in the Matsubara representation~\cite{schrieffer1964theory}, we obtain: 
\begin{eqnarray}
S_{\rm eff}=\int{dR}\Big({\rm {\bar T}r}\ln{G_0^{-1}}-\sum_n^{\infty}\frac{1}{n}{\rm {\bar T}r}\{[\Sigma(R)\tau_3G_0]^n\}\Big)-\int{dR}\Big({\sum_{\bf k}{\eta}p_s^2}+\frac{|\Delta|^2}{J}\Big)+\frac{1}{2}\int{dt}d{\bf q}\frac{|\mu_H(t,{\bf q})|^2}{V_q},
\end{eqnarray}  
where the Green function in the Matsubara representation and momentum space is: 
\begin{equation}\label{GreenfunctionMr}
G_0(p)=\frac{ip_n\tau_0-{\bf p}_s\cdot{\bf v_k}\tau_0+\xi_{\bf k}\tau_3+|\Delta_{\bf k}|\tau_1}{(ip_n-E_{\bf k}^+)(ip_n-E_{\bf k}^-)},  
\end{equation}  
and the self-energy is given by 
\begin{equation}
\Sigma(R)={{\eta}p_s^2}+\mu_H(R)+\frac{\partial_t\delta\theta(R)}{2}.   
\end{equation}
Here, the four-vector momentum $p=(ip_n, {\bf k})$ and the quasiparticle energy $E_{\bm k}^{\pm}={{\bm{p}_s}}\cdot{\bf v_k}\pm\sqrt{\xi_{\bm k}^2+|\Delta_{\bm k}|^2}$.  Keeping the lowest two orders (i.e., $n=1$ and $n=2$) and neglecting the trivial terms under the space-time integral  leads to the result:
\begin{eqnarray}
  S_{\rm eff}&=&\int{dR}\Big\{\sum_{p_n,{\bf k}}\ln[(ip_n-E_{\bf k}^+)(ip_n-E_{\bf k}^-)]-{\tilde \chi}_3{p_s^2}+\chi_{33}\Big[\mu_H(R)+\frac{\partial_t\delta\theta(R)}{2}\Big]^2-\frac{|\Delta|^2}{J}\Big\}+\frac{1}{2}\int{dt}d{\bf q}\frac{|\mu_H(t,{\bf q})|^2}{V_q},
\end{eqnarray}  
where the correlation coefficients are given by
\begin{eqnarray}
{\tilde \chi}_3&=&\sum_{\bf k}\Big[\eta+\sum_{p_n}\eta{\rm Tr}[G_0(p)\tau_3]\Big]=\sum_{\bf k}\Big[\eta+\eta\frac{\xi_{\bf k}}{E_{\bf k}}\Big(f(E_{\bf k}^+)-f(E_{\bf k}^-)\Big)\Big]=-\frac{1}{2}\sum_{\bf k}{\bf v_k^2}\partial_{\xi_{\bf k}}\Big[\frac{\xi_{\bf k}}{2E_{\bf k}}\Big(f(E_{\bf k}^+)-f(E_{\bf k}^-)\Big)\Big], \\
\chi_{33}&=&-\frac{1}{2}\sum_p{\rm Tr}[G_0(p)\tau_3G_0(p)\tau_3]=-\sum_{p_n,{\bf k}}\frac{(ip_n-{\bf p}_s\cdot{\bf v}_{\bf k})^2+\xi_{\bf k}^2-|\Delta_{\bf k}|^2}{(ip_n-E_{\bf k}^+)^2(ip_n-E_{\bf k}^-)^2}=-\sum_{\bf k}\partial_{\xi_{\bf k}}\Big[\frac{\xi_{\bf k}}{2E_{\bf k}}\Big(f(E_{\bf k}^+)-f(E_{\bf k}^-)\Big)\Big].~~~~~~~~~~~\label{chi33}
\end{eqnarray}

Further, through the integration over the Hartree field, one finally arrives at the effective action of the gap and phase fluctuations:
\begin{equation}\label{eff}
 S_{\rm eff}=\int{dR}\Big\{\sum_{p_n,{\bf k}}\ln[(ip_n-E_{\bf k}^+)(ip_n-E_{\bf k}^-)]-{{\tilde \chi}_3p_s^2}-\frac{|\Delta|^2}{J}\Big\}+\int{dt}d{\bf q}D_q\Big(\frac{\partial_{t}\delta\theta_{\bf q}}{2}\Big)^2,
\end{equation}
with $D_q=\frac{\chi_{\rho\rho}}{1+2\chi_{\rho\rho}V_q}$. Here, $\chi_{\rho\rho}=\chi_{33}$ denotes the density-density correlation, and {\small{$V_q=2{\pi}e^2/(q\epsilon_0)$}} for the 2D Coulomb potential.

To facilitate the reading, we would like to clarify and emphasize the reasoning behind our approach.  We begin with the two-point superconducting Green function $G(x,x')$ in Nambu space~\cite{abrikosov2012methods}, expressed in the Matsubara formalism, i.e., the Gorkov Green function. Upon Fourier transforming with respect to the four-vector relative coordinate $r=x-x'$, one arrives at the standard four-momentum representation $(ip_n,{\bf k})$, where ${\bf k}$ is the electronic momentum and $ip_n$ is the fermionic Matsubara frequency. These variables are associated with microscopic quasiparticle dynamics.  On the other hand, the center-of-mass coordinate $R=(x+x')/2$ naturally gives rise to collective (hydrodynamic) variables when treated in the long-wavelength limit. The center-of-mass coordinate transforms into the long-wavelength variables $(i\omega_n,{\bf q})$ where ${\bf q}$ is associated with collective fluctuations and $i\omega_n$ is a bosonic Matsubara frequency, which  are live in a distinct, emergent bosonic Hilbert space. Physically, these two Hilbert spaces  are qualitatively different in nature: one captures short-range fermionic coherence, and the other encodes the slow dynamics of collective bosonic modes such as the Nambu-Goldstone excitation.  It is precisely this separation between fast fermionic (relative) and slow bosonic (center-of-mass) degrees of freedom that underlies the physics of superconducting fluctuations.  In any Green's function formalism, these variables are inherently entangled, leading to a complex mixing of fermionic and bosonic modes across vastly different momentum and frequency scales. Conventional mean-field theories typically neglect such fluctuation dynamics. This simplification allows one to recover the BCS-type solution, but it also completely eliminates the superconducting-phase degrees of freedom and misses the crucial low-energy physics associated with superconducting phase coherence, which becomes particularly important in low-dimensional systems. To systematically incorporate both short-range fermionic and long-range bosonic degrees of freedom, the path-integral formalism is designed to handle such mixing and  allows one to separate slow phase fluctuations from the gapped fermionic quasiparticles by integrating out the fermionic fields.

\subsection{Derivation of thermodynamic equations}

\subsubsection{gap equation}
With the effective action in Eq.~(\ref{eff}), the variation with respect to superconducting gap strength, i.e., $\partial_{|\Delta|}S_{\rm eff}=0$, yields
\begin{eqnarray}
 \frac{|\Delta|}{J}=-\sum_{p_n,{\bf k}}\frac{|\Delta|\zeta_{\bf k}^2}{(ip_n-E^+_{\bf k})(ip_n-E^-_{\bf k})}=-\sum_{{\bf k}}\frac{|\Delta|\zeta_{\bf k}^2}{2E_{\bf k}}[f(E_{\bf k}^+)-f(E_{\bf k}^-)].
\end{eqnarray}
This gap equation is similar to the one in the Fulde-Ferrell-Larkin-Ovchinnikov theory~\cite{fulde1964superconductivity,yang2018fulde,yang21theory}, and in the absence of phase fluctuations recovers the BCS-type gap equation~\cite{abrikosov2012methods}. 

\subsubsection{Nambu–Goldstone phase fluctuations}

The phase-fluctuation field ${\bf p}_s={\bf \nabla_R}\delta\theta({\bf R})/2$ emerges in the superconducting action. Following discussion in Ref.~\cite{benfatto10}, the phase dynamics ${\bf p}_{s}$ can be decomposed into two orthogonal components (Helmholtz decomposition)~\cite{benfatto10}: 
 \begin{equation}
 {\bf p}_s={\bf p}_{s,\parallel}+{\bf p}_{s,\perp},~~\text{with}~~~{\bm \nabla}\times{\bf p}_{s,\parallel}=0~~~\text{and}~~~{\bm \nabla\cdot}{\bf p}_{s,\perp}=0.\label{phased}
\end{equation}
The longitudinal one ${\bf p}_{s,\parallel}$ 
is associated with the gapless Nambu–Goldstone (NG) smooth, long-wavelength phase fluctuations and acts as the superconducting momentum~\cite{ambegaokar61electromagnetic,nambu1960quasi,nambu2009nobel,littlewood81gauge}. Specifically, through the variation $\delta{S_{\rm eff}}$ with respect to the phase fluctuation and considering the long-wavelength approximation, in the frequency and momentum space, the equation of motion of the NG phase mode reads
\begin{equation}
\Big(2f_sq^2-D_q\omega^2\Big)\frac{\delta\theta(\omega,{\bf q})}{2}=0,  \label{PEF}  
\end{equation}
where the superconducting phase stiffness is determined by 
\begin{eqnarray}
f_s&=&\frac{\tilde \chi_3}{2}\!+\!\frac{\partial_{\bf p_s}}{4}\Big[{\sum_{p_n{\bf k}}}\frac{2{\bf v_k}(ip_n\!-\!{\bf v_k}\!\cdot\!{\bf p}_s)}{(ip_n\!-\!E_{\bf k}^+)(ip_n\!-\!E_{\bf k}^-)}\Big]={\sum_{\bf k}}\frac{\mathrm{v}^2_{\bf k}}{4}\Big\{-\partial_{\xi_{\bf k}}\Big[\frac{\xi_{\bf k}}{E_{\bf k}}\Big(f(E_{\bf k}^+)-f(E_{\bf k}^-)\Big)\Big]+\partial_{E_{\bf k}}\Big[\frac{f(E_{\bf k}^+)-f(E_{\bf k}^-)}{2}\Big]\Big\}\nonumber\\
  &=&\sum_{\bf k}\mathrm{v}^2_{\bf k}\frac{|\Delta_{\bf k}|^2}{4E_{\bf k}}\partial_{E_{\bf k}}\Big[\frac{f(E_{\bf k}^+)-f(E_{\bf k}^-)}{2E_{\bf k}}\Big]\approx\sum_{\bf k}\mathrm{v}^2_{\bf k}|\Delta_{\bf k}|^2\frac{f(E_{\bf k}^-)-f(E_{\bf k}^+)}{8E_{\bf k}^3}. \label{scfss}
\end{eqnarray}
As a self-consistent check, $f_s$ for a parabolic normal-state energy dispersion
reduces to the conventional superconducting phase stiffness $\hbar^2n_s/(4m^*)$, with $m^*$ being 
the normal-state effective mass and $n_s=2E_F\sum_{\bf k}|\Delta_{\bf k}|^2\frac{f(E_{\bf k}^-)-f(E_{\bf k}^+)}{2E_{\bf k}^3}$ denoting the superfluid density. In this case, Eq.~(\ref{PEF}) is the same as the equation of motion of the NG mode in the literature~\cite{anderson63plasmons,ambegaokar61electromagnetic,littlewood81gauge,nambu1960quasi,yang21theory,yang19gauge,sun20collective}, and the expression of the superfluid density here is also the same as that obtained in the literature by various approaches~\cite{yang19gauge,sun20collective,yang21theory,yang24thermodynamic}. 
It should be emphasized that the above derivation holds in the clean limit. In practice, disorder is unavoidable in layered systems and affects the amplitude (gap) and phase (stiffness) sectors differently. The superfluid density, being a current–current correlation~\cite{PhysRevB.99.224511,PhysRevB.106.144509}, is reduced by the elastic scattering, as realized by various theoretical approaches using Gorkov theory~\cite{abrikosov2012methods}, Eilenberger transport~\cite{eilenberger1968transformation,PhysRevLett.25.507,PhysRevB.99.224511}, gauge-invariant kinetic equations~\cite{PhysRevB.98.094507}, and diagrammatic formulations incorporating {\sl vertex} corrections to the current–current correlation~\cite{PhysRevB.99.224511,PhysRevB.106.144509}. While illuminating, such microscopic treatments are cumbersome for practical modeling. Here we directly adopt Tinkham's interpolation in the context of $s$-wave superconductors~\cite{tinkham2004introduction}, 
 which approximate the penetration depth $\lambda$ as 
 $\lambda^2=\lambda_{\rm clean}^2(1+\xi/l)$ and hence $f_s\rightarrow f_{s,{\rm clean}}/(1+\xi/l)$. Thus, the superfluid-density expression [Eq.~(\ref{fss})] in our framework becomes 
 \begin{equation}\label{SD2}
f_s=\frac{1}{1+\xi/l}\sum_{\bf k}\mathrm{v}^2_{\bf k}|\Delta_{\bf k}|^2\frac{f(E_{\bf k}^-)-f(E_{\bf k}^+)}{8E_{\bf k}^3},  
\end{equation}
 where $\xi=\hbar \mathrm{v}_F/|\Delta|$ is the superconducting coherence length and $l=\mathrm{v}_F/\Gamma$ is the mean free path, with $\Gamma$ representing the effective  scattering rate, encompassing the quasiparticle decay and transport processes. This prescription is consistent with the Mattis–Bardeen dirty-limit result~\cite{PhysRevB.109.064508,PhysRev.111.412} and Nam’s extension to arbitrary scattering strength~\cite{nam1967theory}.\\

The statistical average of NG phase fluctuations must be treated within quantum-statistical framework, as a consequence of the NG bosons dictated by fundamental NG  theorem~\cite{nambu1960quasi,nambu2009nobel}, following the spontaneous breaking of $U(1)$ symmetry in superconductors.  From the equation of motion of NG mode, one can introduce a thermal field and then calculate the thermal phase fluctuation via the fluctuation dissipation theorem, or directly calculate the thermal phase fluctuation through the Bosonic Green function within Matsubara representation. The two methods are equivalent and lead to the same result: 
 $\langle{\bf p}_{s,\parallel}\rangle=0$ and
\begin{equation}\label{BET}
\langle{p_{s,\parallel}^2}\rangle=S_{\rm th}(T)+S_{\rm zo},
\end{equation}
with the thermal-excitation part
\begin{equation}\label{th}
S_{\rm th}(T)=\int\frac{d{\bf q}}{(2\pi)^2}\frac{q^2n_B(\omega_{\rm NG})}{D_{q}\omega_{\rm NG}(q)}, 
\end{equation}
and the zero-point oscillation part 
\begin{equation}\label{zo}
S_{\rm zo}=\int\frac{d{\bf q}}{(2\pi)^2}\frac{q^2}{2D_{q}\omega_{\rm NG}(q)}.  
\end{equation}
Here, $n_B(x)$ denotes the Bose-Einstein distribution.
 This part enters the superconducting gap equation and is expected to renormalize the superconducting  gap~\cite{PhysRevB.97.054510,yang21theory,PhysRevB.70.214531,PhysRevB.102.060501}, in a gauge manner analogous to the way a vector potential can influence the superconducting  gap~\cite{ambegaokar61electromagnetic,nambu1960quasi,nambu2009nobel}. We present the detailed derivation of this excitation in the following. 
 
{\sl Fluctuation dissipation theorem.---}Considering the thermal fluctuations, the dynamics of the phase from Eq.~(\ref{PEF}) is given by
\begin{equation}
\Big(2f_sq^2-D_q\omega^2+i\omega{\gamma}\Big)\frac{\delta\theta({\omega,{\bf q}})}{2}={J}_{\rm th}(\omega,{\bf q}).    
\end{equation}  
Here, we have introduced a thermal field ${\bf J}_{\rm th}(\omega,{\bf q})$ that obeys the fluctuation-dissipation theorem~\cite{Landaubook}:
\begin{equation}
\langle{J}_{\rm th}(\omega,{\bf q}){J}^*_{\rm th}(\omega',{\bf q}')\rangle=\frac{(2\pi)^3\gamma\omega\delta(\omega-\omega')\delta({\bf q-q'})}{\tanh(\beta\omega/2)},   
\end{equation}
and $\gamma=0^+$ is a damping constant. From this dynamics, the statistic average of NG phase fluctuations is given by
\begin{eqnarray}
  \langle{p_{s,\parallel}^2}\rangle&=&\int\frac{d\omega{d\omega'}d{\bf q}d{\bf q'}}{(2\pi)^6}\frac{({\bf q}\cdot{\bf q'})\langle{J}_{\rm th}(\omega,{\bf q}){J}^*_{\rm th}(\omega',{\bf q}')\rangle}{\big[D_q\big(\omega^2\!-\!\omega^2_{\rm NG}({\bf q})\big)\!-\!i\omega\gamma\big]\big[D_{q'}({\omega'}^2\!-\!\omega^2_{\rm NG}({\bf q'})\big)\!+\!i\omega'\gamma\big]}=\int\frac{d\omega{d{\bf q}}}{(2\pi)^3}\frac{q^2\gamma\omega/\tanh(\beta\omega/2)}{\big[D_q\big(\omega^2\!-\!\omega^2_{\rm NG}({\bf q})\big)\big]^2+\omega^2\gamma^2}\nonumber\\
&=&\int\frac{{d{\bf q}}}{(2\pi)^2}\frac{q^2}{2D_q\omega_{\rm NG}(q)}[2n_B(\omega_{\rm NG})+1].\label{EEEE}
\end{eqnarray}
where {\small{$\omega_{\rm NG}(q)=\sqrt{2f_sq^2/D_q}$}} is the energy spectrum of the phase mode. \\

{\sl Matsubara formalism.---}Within the Matsubara formalism, by mapping into the imaginary-time space, the statistic average of the NG phase fluctuation reads~\cite{abrikosov2012methods}
\begin{eqnarray}
  \langle{p_{s,\parallel}^2}\rangle&\!=\!&\int\frac{d{\bf q}}{(2\pi)^2}q^2\Big[\Big\langle\Big|\frac{\delta\theta^*(\tau,{\bf q})}{2}\frac{\delta\theta(\tau,{\bf q})}{2}e^{-\int^{\beta}_{0}d\tau{d{\bf q}}D_q{\delta\theta^*(\tau,{\bf q})}(\omega_{\rm NG}^2-\partial_{\tau}^2){\delta\theta(\tau,{\bf q})}/4}\Big|\Big\rangle\Big]\nonumber\\
  &\!=\!&\int\frac{d{\bf q}}{(2\pi)^2}q^2\Big[\frac{1}{\mathcal{Z}_0}{\int}D\delta\theta{D\delta\theta^*}\frac{\delta\theta^*(\tau,{\bf q})}{2}\frac{\delta\theta(\tau,{\bf q})}{2}e^{-\int^{\beta}_{0}d\tau{d{\bf q}}D_q{\delta\theta^*(\tau,{\bf q})}(\omega_{\rm NG}^2-\partial_{\tau}^2){\delta\theta(\tau,{\bf q})}/4}\Big]\nonumber\\
   &=&\!\!\!\int\frac{d{\bf q}}{(2\pi)^2}q^2\frac{1}{\mathcal{Z}_0}{\int}D\delta\theta{D\delta\theta^*}\delta_{J^*_{\bf q}}\delta_{J_{\bf q}}e^{-\int^{\beta}_{0}d\tau{d{\bf q}}[D_q{\delta\theta^*(\tau,{\bf q})}(\omega_{\rm NG}^2-\partial_{\tau}^2){\delta\theta(\tau,{\bf q})}/4+J_{\bf q}\delta\theta({\bf q})/2+J^*_{\bf q}\delta\theta^*({\bf q})/2]}\Big|_{J=J^*=0}\nonumber\\
  &=&\!\!\!\int\frac{d{\bf q}}{(2\pi)^2}\frac{q^2}{D_q}\delta_{J^*_{\bf q}}\delta_{J_{\bf q}}\exp\Big\{-\int_0^{\beta}{d\tau}\sum_{\bf q'}J_{\bf q'}\frac{1}{\partial_{\tau}^2-\omega_{\rm NG}^2}J^*_{\bf q'}\Big\}\Big|_{J=J^*=0}=-\int\frac{d{\bf q}}{(2\pi)^2}\frac{1}{\beta}\sum_{\omega_n}\frac{q^2}{D_q}\frac{1}{(i\omega_n)^2\!-\!\omega_{\rm NG}^2}\nonumber\\
 &=&\int\frac{{d{\bf q}}}{(2\pi)^2}\frac{q^2}{2D_q\omega_{\rm NG}(q)}[2n_B(\omega_{\rm NG})+1],
\end{eqnarray}
which is the same as the one in Eq.~(\ref{EEEE}) obtained via fluctuation dissipation theorem. Here, $\omega_n=2n\pi{T}$ represents the Bosonic Matsubara frequencies; ${J_{\bf q}}$ denotes the generating functional and $\delta{J_{\bf q}}$ stands for the functional derivative.  \\

It should be emphasized that Eq.~(\ref{BET}) determines only the magnitude of the NG phase fluctuations. In realistic systems,  these fluctuations occur across all directions as a result of the continuous symmetry of the superconducting phase. A proper  treatment must incorporate the statistical average over all possible fluctuation directions to accurately describe physical observables. Therefore, one must perform a statistical average over NG phase fluctuations in all directions when evaluating the gap equation, the density-density correlation function, and the superconducting phase stiffness---quantities that are inherently affected by NG superconducting phase fluctuations, as introduced in the main text.  This reflects the physical scenario that while no net supercurrent arises in the model, i.e., $\langle{\bf p}_{s,\parallel}\rangle=0$, the impact of the long-range NG phase fluctuations is incorporated through finite $\langle{p}^2_{s,\parallel}\rangle\ne0$ to the key superconducting properties. 

\subsubsection{Berezinskii–Kosterlitz–Thouless fluctuations}

The transverse  component ${\bf p}_{s,\perp}$ in Eq.~(\ref{phased}) encodes Berezinskii–Kosterlitz–Thouless (BKT) fluctuations, since only this part carries vorticity: 
\begin{equation}
\pi\sum_jq_j=\oint{\bf p}_s\cdot{d{\bf l}}=\int_S({\bm \nabla}\times{\bf p_s})\cdot{d{\bf s}}=\int_S({\bm \nabla}\times{\bf p_{s,\perp}})\cdot{d{\bf s}},
\end{equation}
or equivalently,
\begin{equation}
{\bm \nabla}\times{\bf p_{s,\perp}}=\pi\sum_jq_j\delta({\bf r-r_j}),
\end{equation}
which represents a set of point-like topological defects (vortices and antivortices) with integer vorticity $q_j\in\mathbb{Z}$ (topological charge). These topological defects disorder the phase while leaving the pairing amplitude $|\Delta|$ essentially intact, except within vortex cores. Their effect is to renormalize the bare superfluid stiffness $f_s$ [Eq.~(\ref{SD2})], which enters the well-established BKT renormalization-group (RG) equations~\cite{benfatto10,PhysRevB.110.144518,PhysRevB.80.214506,PhysRevB.77.100506,PhysRevB.87.184505}:
\begin{equation}\label{BKT1}
\frac{dK}{dl}=-K^2g^2~~~\text{and}~~~\frac{dg}{dl}=(2-K)g,
\end{equation}
with initial conditions of the dimensionless stiffness $K(l)$ and the vortex fugacity $g(l)$: 
\begin{equation}\label{BKT0}
K(l=0)=\frac{\pi{f_s}}{k_BT}~~~\text{and}~~~g(l=0)=2\pi{e}^{-\mu_v(T)/(k_BT)},
\end{equation}
where $\mu_v$ is the vortex–core energy. Following Ref.~\cite{benfatto10}, in the 2D limit $\mu_v$ can be estimated as the condensation energy lost inside a core of radius $L_v=\hbar v_F/(\pi|\Delta|)$, yielding $\mu_v=\pi{L^2_v}E_{c}\frac{n_s}{n}\approx\frac{2}{\pi}f_s$, with $E_c=D|\Delta|^2/2$ the condensation-energy density. Integrating the RG flow to $l\to\infty$ yields the renormalized stiffness
\begin{equation}\label{BKT2}
{\bar f}_s=\frac{k_BT}{\pi}K(l=\infty),
\end{equation}
accounting for vortex–antivortex fluctuations. The separatrix $2-\pi K=0$ corresponds to the Nelson–Kosterlitz universal jump~\cite{benfatto10,PhysRevB.110.144518,PhysRevB.80.214506,PhysRevB.77.100506,PhysRevB.87.184505}, with flows below it running to the disordered phase (free vortices) and flows above it to a finite $K$ (bound vortex pairs). Thus, in the longitudinal–transverse decomposition, only the curl-free longitudinal part (NG phase fluctuations) enters the gap equation and drives uniform pair breaking, while the transverse component contributes exclusively through BKT physics, renormalizing the superfluid stiffness but leaving the homogeneous gap equation unaffected.

\subsubsection{Energy–diagonal (on-shell) projection}

In strongly correlated systems the single-particle excitations are no longer sharp Bloch states characterized solely by the bare dispersion $\xi_{\bm k}$, but instead exhibit substantial incoherence and spectral broadening and is associated with electronic correlations. As a result, the low-energy electronic properties are encoded in the fully interacting spectral function $A({\bm k},\omega)$, which contains both coherent quasiparticle peaks and incoherent background weight as well as strong electronic correlations~\cite{worm23numerical,worm24fermi}.  In the Cooper channel, the pairing kernel formally involves a two-frequency convolution of $A({\bm k},\omega)$ and $A(-{\bm k},\omega')$~\cite{abrikosov2012methods}.
 In practice, evaluating the full two-frequency kernel of the gap equation, and in particular, of the superfluid stiffness is generally intractable. A controlled simplification is obtained by invoking the energy-diagonal (on-shell) projection: since pairing is dominated by nearly on-shell processes~\cite{PhysRevB.57.R11093,PhysRevB.64.075116,RevModPhys.78.17} where two fermions at opposite momenta and nearly the same energy form a Cooper pair, the main contribution arises from the diagonal sector $\omega\simeq\omega'$. This allows one to replace the bare dispersion by an energy average weighted with the interacting spectral function (spectral representation)~\cite{mahan2013many,PhysRevB.57.R11093}, 
\begin{equation}\label{EDPJ}
\sum_{\bf k}F(\xi_{\bf k})\longrightarrow\sum_{\bf k}\int\frac{d\omega}{2\pi}A({\bf k},\omega)F(\omega). 
\end{equation}
The approximation is exact in the clean limit $A({\bm k},\omega)=2\pi\delta(\omega-\xi_{\bm k})$, and remains accurate whenever the interacting correlated normal-state spectral function $A({\bm k},\omega)$ retains a well-defined quasiparticle peak, as considered in the present study and in numerical investigations of Hubbard-model-based theories~\cite{worm24fermi,worm23numerical}. In this way,  correlation and scattering effects are incorporated through the redistribution of spectral weight, while the algebraic structure of the superconducting kernel is preserved.

\section{Single-particle excitation spectrum of the solvable model}
\label{appC}

The solvable correlated model used for the normal-state single-particle excitation spectrum in the main text was originally introduced by Worm {\sl et al.}~\cite{worm24fermi}, where this model was thoroughly justified by comparison with numerically demanding many-body calculations based on the full local Hubbard model, using the dynamical vertex approximation (D$\Gamma$A) method. For completeness and to facilitate the presentation, we summarize the model here. We begin with the full local Hubbard interaction and its Fourier transform:
\begin{equation}
\hat{H}_i=\frac{U}{2}\sum_{\sigma{i}}{\hat n}_{\sigma{i}}{\hat n}_{-\sigma{i}}=\frac{U}{2}\sum_{\bf kk'q}c^{\dagger}_{\sigma{\bf k}}c_{\sigma{\bf k'}}c^{\dagger}_{-\sigma{\bf k'+q}}c_{-\sigma{\bf k+q}}.
\end{equation}
Enforcing translational symmetry on the ${\bf k}$-space density operator by imposing the constraint $\delta_{\mathbf{k}, \mathbf{k}'}$ simplifies the interaction structure, while still incorporating strong correlation effects. Such a procedure allows one to isolate the most relevant correlations that are responsible for the {\sl collective} phenomena and macroscopic properties. Then, one has 
\begin{equation}\label{HU}
\hat{H}_i=\frac{U}{2}\sum_{\bf k}{\hat n}_{\sigma{\bf k}}{\hat n}_{-\sigma{\bf k}}+\frac{U}{2}\sum_{\bf k}{\hat n}_{\sigma{\bf k}}{\hat n}_{-\sigma{\bf k+Q}}+\frac{U}{2}\sum_{\bf q\ne{Q},0}\sum_{\bf k}c^{\dagger}_{\sigma{\bf k}}c_{\sigma{\bf k}}c^{\dagger}_{-\sigma{\bf k+q}}c_{-\sigma{\bf k+q}},
\end{equation}
where we have singled out the terms corresponding to $\mathbf{q} = 0$ and $\mathbf{q} = \mathbf{Q} = (\pi, \pi)$ from the summation. 

Notably, keeping only the first term, which restricts the interaction to zero momentum transfer ($\mathbf{q} = 0$), results in the interaction in the Hatsugai-Kohmoto model, which describes a Mott insulator that exhibits non-Fermi liquid behavior upon doping. Keeping only the second term, which restricts the interaction to the momentum transfer $\mathbf{q} = \mathbf{Q} = (\pi, \pi)$, simplifies to the interaction model proposed in Ref.~\cite{worm24fermi},
\begin{equation}\label{Hamiltonian}
    \hat{H}_i= \frac{{\mathcal{V}}}{2}\hat{n}_{\bm{k} \sigma} \hat{n}_{\bm{k} + \bm{Q} \bar{\sigma}}.   
\end{equation}
Then, the correlated normal-state model Hamiltonian becomes nearly diagonal, $\hat{H} = \sum_{\bm{k}\in \mathrm{HBZ}} \sum_\sigma \hat{H}_{\bm{k} \sigma}$, with HBZ denoting half of Brillouin zone (related to the other half via $\bm{k} \rightarrow \bm{k} + \bm{Q}$), and the momentum-resolved component is
\begin{equation}
    \hat{H}_{\bm{k}\sigma} = \epsilon_{\bm{k}} \hat{n}_{\bm{k}\sigma} + \epsilon_{\bm{k}+\bm{Q}} \hat{n}_{\bm{k}+\bm{Q} \bar{\sigma}} + {\mathcal{V}}\hat{n}_{\bm{k}\sigma} \hat{n}_{\bm{k}+\bm{Q}\bar{\sigma}}.
\end{equation}
This structure implies that for each $\bm{k}$ and spin $\sigma$, only two single-particle states are involved: $\bm{k} \sigma$ and $\bm{k} + \bm{Q} \bar{\sigma}$. Denoting the occupation numbers of the $\bm{k} \sigma$ and $\bm{k}+\bm{Q} \bar{\sigma}$ states by $a$ and $b$ in the ket $|a,b\rangle$  respectively, the full Fock space consists of four states: $|0,0\rangle$, $|1,0\rangle$, $|0,1\rangle$, and $|1,1\rangle$. In this basis, $\hat{H}_{\bm{k}\sigma}$ is diagonal and spin-independent: 
\begin{equation}
    \hat{H}_{\bm{k}\sigma} = \mathrm{diag}\{ 0, \epsilon_{\bm{k}}, \epsilon_{\bm{k}+\bm{Q}}, \epsilon_{\bm{k}} + \epsilon_{\bm{k}+\bm{Q}} + {\mathcal{V}} \}.
\end{equation}
Then, one can use the Lehmann representation to calculate the single-particle excitation spectrum function~\cite{mahan2013many},
\begin{equation}
    A_\sigma(\bm{k},\omega) = \sum_{n n'} | \langle n | \hat{c}_{\bm{k}\sigma} | n' \rangle |^2 \frac{e^{-\beta E_n} + e^{-\beta E_{n'}}}{Z} \delta(\omega + E_n - E_{n'}), \label{eq:lehm}
\end{equation}
where $|n\rangle$ and $E_n$ are the eigenstates and eigenenergies, and $Z = \sum_n e^{-\beta E_n}$ is the partition function. 
Substituting the four Fock states and their energies into Eq.~(\ref{eq:lehm}) and summing over allowed transitions, we obtain the spectral function (dropping the spin index for clarity): 
\begin{equation}\label{ASPF}
    A(\bm{k},\omega) = (1 - n_{\bm{k}+\bm{Q}}) \delta(\omega - \epsilon_{\bm{k}}) + n_{\bm{k}+\bm{Q}} \delta(\omega - \epsilon_{\bm{k}} - {\mathcal{V}}),
\end{equation}
where the thermal occupation factor $n_{\bm{k}+\bm{Q}}$ is given by
\begin{equation}
n_{\bm{k}+\bm{Q}}=\frac{e^{-\beta\epsilon_{{\bm{k}+\bm{Q}}}}+e^{-\beta(\epsilon_{\bm{k}}+\epsilon_{{\bm{k}+\bm{Q}}}+{\mathcal{V}})}}{1+e^{-\beta\epsilon_{\bm{k}}}+e^{-\beta\epsilon_{{\bm{k}+\bm{Q}}}}+e^{-\beta(\epsilon_{\bm{k}}+\epsilon_{{\bm{k}+\bm{Q}}}+{\mathcal{V}})}}.
\end{equation}
This result clearly illustrates the two-peak structure of the spectral function, originating from the correlation effect related to the presence or absence of an electron at $\bm{k}+\bm{Q}$.  This restriction captures the specific coupling associated to the antiferromagnetic (AF) spin fluctuations, motivated by the fact that AF spin fluctuations dominate the leading low-energy interaction channel of cuprates.  Ref.~\cite{worm24fermi} has demonstrated that this simplified interaction reproduces the essential features of the original local Hubbard model as obtained via numerically demanding many-body calculations (by means of cutting-edge D$\Gamma$A quantum many-body scheme), particularly in the doping range away from the parent insulating phase.  These include the doping evolution of the Fermi surface and the momentum-resolved spectral function.  Thus, this simplified interaction model offers a controlled and physically motivated reduction that can effectively capture the relevant electronic correlations and the doping evolution of the Fermi surface and the
momentum-resolved spectral function. 

{The effective interaction strength here is denoted by $\mathcal V$  rather than $U$. Physically, from a renormalization-group perspective, the interaction with ${\bf Q}=(\pi,\pi)$  captures a specific low-energy interaction channel associated with antiferromagnetic spin fluctuations, which are widely believed to dominate the leading low-energy correlations in cuprates. As such, ${\mathcal{V}}$ should be viewed as an effective coupling corresponding to a distinct low-energy fixed point, obtained after integrating out high-energy degrees of freedom (e.g., interaction channels at other ${\bf Q}\ne(\pi,\pi)$). Within such a picture, a substantial reduction in magnitude relative to the bare on-site interaction $U$ can be expected, consistent with the numerical calculation within the cutting-edge D$\Gamma$A scheme.}

\section{Simulation treatment}

To visualize the spectral function $A(\bm{k}, \omega)$ in Fig.~1 of the main text, we approximate the $\delta$-functions by Lorentzian profiles with half width at half maximum $\Gamma_A = 0.04t$, which yields a smooth representation suitable for graphical presentation.

In the actual simulations, however, we employ a narrower broadening $\Gamma_A = 0.01t$ in order to resolve fine spectral features near the Fermi level more accurately. For the momentum ${\bf k}$ summation, we sample the entire Brillouin zone. For the superconducting gap equation, we introduce a frequency cutoff by multiplying the integral kernel with the factor $1/[1+(\omega/\omega_D)^2]$, where $\omega_D = t/3$. This smooth cutoff has a clear physical origin: the effective pairing interaction in correlated superconductors is not instantaneous but mediated by bosonic modes (e.g., spin or phonon fluctuations) that are characterized by a finite frequency scale $\omega_D$. Consequently, the attractive interaction is effective only within an energy window of order $\omega_D$ around the Fermi level, and is strongly suppressed at higher frequencies. For the integrals associated with bosonic excitations, following the conventional treatment~\cite{yang21theory,PhysRevB.97.054510,PhysRevB.70.214531}, we impose a momentum cutoff $q_{\rm cutoff} = 1/\xi_0$, where $\xi_0 = \hbar v_F/|\Delta(T=0)|$ is the superconducting coherence length at zero temperature. The effective scattering rate for the disorder modification in the superfluid stiffness, encompassing quasiparticle decay and transport processes, is set to $\Gamma = 0.02t$.

A subtle point in the specific simulation arises in the treatment of zero-point oscillations of the  NG excitations. The phase mode described in Eq.~(\ref{BET}) consists of two contributions: thermal excitations and quantum zero-point oscillations. Even at $T=0$, the latter yields a finite value of the phase fluctuations, reflecting the intrinsic quantum nature of the superconducting phase. To account for these quantum fluctuations, we adopt a field-theoretic perspective~\cite{peskin2018introduction}, where the vacuum is not empty but populated by zero-point excitations enforced by the uncertainty principle, and the vacuum-related, temperature-independent contribution is absorbed into the ground-state parameters.  This issue is not unique to superconductivity; it is well known in {\sl ab initio} studies of lattice vibrations within density functional theory. There, the zero-point contributions are usually absorbed into renormalized model parameters~\cite{verdi2023quantum,wu2022large}, such that subsequent phonon excitations are governed primarily by the Bose distribution $n_B(x)$. Analogously, in our framework we absorb the zero-point contributions into a redefinition of the pairing interaction strength. Concretely, starting from the bare pairing potential $J$, we determine the actual zero-temperature superconducting gap $|\Delta(T=0)|$ by numerically solving the full set of coupled equations self-consistently: the gap equation,
\begin{equation}\label{DTZ}
\frac{|\Delta(T=0)|}{J}=-{F[p_{s,\parallel}^2=S_{\rm zo},|\Delta(T=0)|,T=0]},
\end{equation}
together with the phase fluctuation amplitude, the density–density correlation function, and the phase stiffness at $T=0$. We then define a renormalized pairing strength $\tilde J$ such that the zero-point effects are fully absorbed into the ground-state properties: 
\begin{equation}\label{RPP}
\frac{|\Delta(T=0)|}{\tilde J}=-{F[p_{s,\parallel}^2=0,|\Delta(T=0)|,T=0]}. 
\end{equation}
Then, the renormalized gap equation at finite temperature takes the form:
\begin{equation}\label{RGE}
\frac{|\Delta(T)|}{\tilde J}=-{F[p_{s,\parallel}^2=S_{\rm th},|\Delta(T)|,T]},
\end{equation}
This procedure guarantees that the zero-temperature superconducting gap is preserved. 

In summary, for each doping level (Fermi energy), our simulation proceeds in two steps. First, within the energy–diagonal (on-shell) projection [Eq.~(\ref{EDPJ})] and the  single-particle excitation spectrum function [Eq.~(\ref{ASPF})],  using an iterative procedure, we self-consistently solve the zero-temperature gap equation [Eq.~(\ref{DTZ})], the zero-point NG phase fluctuations [Eq.~(\ref{zo})],   the density–density correlation function [Eq.~(\ref{chi33})], and the phase stiffness at $T=0$ [Eq.~(\ref{SD2})],  thereby obtaining the zero-point–renormalized gap $|\Delta(T=0)|$ and hence the pairing potential [Eq.~(\ref{RPP})]. Next, we solve the finite-temperature gap equation [Eq.~(\ref{RGE})], the thermal NG phase
fluctuations [Eq.~\ref{th}], the density–density correlation function [Eq.~(\ref{chi33})], and the phase stiffness at $T\ne0$ [Eq.~(\ref{SD2})] in the same iterative manner  within the energy–diagonal (on-shell) projection [Eq.~(\ref{EDPJ})] and the  single-particle excitation spectrum function [Eq.~(\ref{ASPF})]. Substituting the resulting bare finite-temperature superfluid stifness into the BKT renormalization-group equations [Eqs.~(\ref{BKT1})-(\ref{BKT2})] then allows
us to determine the full set of superconducting properties. The superconducting transition temperature $T_c$ follows from $\bar f_s(T_c)=0$, while the gap-closing temperature [$\Delta(T=T_{\rm os})=0$] is denoted $T_{\rm os}$.

\end{widetext}

\end{appendix}

%\bibliography{sc-refs}% Produces the bibliography via BibTeX.
\end{document}